\documentclass[fleqn,usenatbib]{mnras}

\usepackage[T1]{fontenc}
\usepackage{ae,aecompl}

\usepackage{graphicx}	
\usepackage{amsmath}	
\usepackage{amssymb}	
\usepackage{soul}	
\usepackage{color,xcolor}	
\usepackage{pdflscape}
\usepackage{soul} 

\title[Properties of $\gamma$-ray quiet FSRQs]{The physical properties of $\gamma$-ray quiet flat-spectrum radio quasars: why are they undetected by $Fermi$-LAT?}

\author[Xue-Jiao Deng et al.]{\parbox{\linewidth}{
Xue-Jiao Deng,$^{1}$
Rui Xue,$^{2}$\thanks{E-mail: ruixue@zjnu.edu.cn}
Ze-Rui Wang,$^{3,4}$\thanks{E-mail: zerui\_wang62@163.com}
Shao-Qiang Xi,$^{5}$
Hu-Bing Xiao,$^{6}$
Lei-Ming Du $^{1}$\\
and Zhao-Hua Xie $^{1}$}
\\
\\
$^{1}$Department of Physics, Yunnan Normal University, 650500, Kunming, People's Republic of China\\
$^{2}$Department of Physics, Zhejiang Normal University, Jinhua 321004, People's Republic of China\\
$^{3}$School of Astronomy and Space Science, Nanjing University, Nanjing 0110093, People's Republic of China\\
$^{4}$Key laboratory of Modern Astronomy and Astrophysics (Nanjing University), Ministry of Education, Nanjing 210023, \\
People's Republic of China\\
$^{5}$Key Laboratory of Particle Astrophysics \& Experimental Physics Division \& Computing Center, Institute of High Energy Physics, \\ 
Chinese Academy of Sciences, 100049 Beijing, China\\
$^{6}$Key Lab for Astrophysics, Shanghai Normal University, Shanghai 200234, People's Republic of China\\
}

\date{Accepted 2021 July 19. Received 2021 June 27; in original form 2021 April 19.}

\pubyear{2021}

\begin{document}
\label{firstpage}
\pagerange{\pageref{firstpage}--\pageref{lastpage}}
\maketitle
\begin{abstract}
During a decade of the $Fermi$-Large Area Telescope (LAT) operation, thousands of blazars have been detected in the $\gamma$-ray band. However, there are still numbers of blazars that have not been detected in the $\gamma$-ray band. In this work, we focus on investigating why some flat-spectrum radio quasars (FSRQs) are undetected by $Fermi$-LAT. By cross-matching the Candidate Gamma-ray Blazars Survey catalog with the Fourth Catalog of Active Galactic Nuclei Detected by the $Fermi$-LAT, we select 11 $\gamma$-ray undetected ($\gamma$-ray quiet) FSRQs as our sample whose quasi-simultaneous multi-wavelength data are collected. In the framework of the conventional one-zone leptonic model, we investigate their underlying physical properties and study the possibility that they are undetected with $\gamma$-ray by modeling their quasi-simultaneous spectral energy distributions. In contrast to a smaller bulk Lorentz factor suggested by previous works, our results suggest that the dissipation region located relatively far away from the central super-massive black hole is more likely to be the cause of some $\gamma$-ray quiet FSRQs being undetected by $Fermi$-LAT.
\end{abstract}




\section{Introduction}

Blazars are a class of highly variable jet-dominated active galactic nuclei (AGN) whose relativistic jets point close to our line of sight \citep{1978PhyS...17..265B}. The population of blazars consists of flat-spectrum radio quasars (FSRQs) with strong emission lines (equivalent width, $EW > 5 $~\AA) and BL Lacertae objects (BL Lacs) showing weak or no emission lines \citep{1995PASP..107..803U}. The broadband spectral energy distributions (SEDs) of blazars generally display two non-thermal emission humps. The low-energy hump, from radio to optical/X-ray, is agreed to be synchrotron emission from relativistic electrons in the dissipation regions located inside the jet. In leptonic scenarios, the high-energy hump, from X-ray to $\gamma$-ray, is believed to be the inverse Compton (IC) emission from relativistic electrons that up-scatter either the synchrotron photons emitted by the same population of relativistic electrons \citep[synchrotron self-Compton, SSC;][]{1985ApJ...298..114M}, or external photons (external Compton, EC) from an accretion disk \citep{1993ApJ...416..458D}, 
a broad-line region \citep[BLR,][]{1994ApJ...421..153S}, and a dusty torus \citep[DT,][]{2000ApJ...545..107B}. In addition, the hadronic emission, such as proton synchrotron radiation and emission from secondary particles \citep[e.g.,][]{2000NewA....5..377A}, could be another possible origin for the high-energy hump, and the accompanying high-energy neutrino emission might be observed simultaneously \citep{2018Sci...361..147I, 2018Sci...361.1378I}.

The $Fermi$-Large Area Telescope ($Fermi$-LAT) finds that the diffuse extragalatic $\gamma$-ray background is significantly dominated by emission from blazars \citep{2015ApJ...800L..27A, 2016PhRvL.116o1105A}. However, there are still plenty of blazars that have not been detected in the $\gamma$-ray band (hereafter the $\gamma$-ray quiet blazars). Various explanations have been proposed. Radio observations suggest that the jets of $\gamma$-ray quiet blazars are oriented at preferentially larger angles to the line of sight resulting in a weaker beaming effect \citep{2009A&A...507L..33P}. In addition, some studies find that $\gamma$-ray quiet blazars have smaller jet Doppler factor \citep[e.g.,][]{2014A&A...562A..64W, 2015ApJ...810L...9L, 2017ApJ...851...33P}. \citet{2015ApJ...810L...9L} finds that blazars with the low-energy humps peaked below 10$^{13.4}$~Hz tend to have high-energy humps peaked below the 0.1~GeV threshold of $Fermi$-LAT, and are thus less likely to be detected by the $Fermi$-LAT. By modeling the broadband SEDs of a large sample of $\gamma$-ray quiet blazars, \citet{2017ApJ...851...33P} suggests that the $\gamma$-ray quiet blazars have smaller break Lorentz factors $\gamma_{\rm b}$ of electron energy distributions compared to that of the $\gamma$-ray loud blazars, which makes the high-energy humps peak mostly in the MeV band instead of the GeV band, which is harder to be detected by $Fermi$-LAT. On the other hand, modeling results suggest that $\gamma$-ray quiet blazars have larger spectral indices $q$ after $\gamma_{\rm b}$, which make the IC spectrum steeper and thus avoid detection by $Fermi$-LAT. However, no quasi-simultaneous data are applied in their modeling. Even if only historical data are available, $\gamma_{\rm b}$ and $q$ are still not constrained well and set arbitrarily. 

It is well known that the properties of blazars' jet radiation often change dramatically. Not only does the flux in each band vary greatly, but the peak frequencies of the two humps may also shift significantly \citep[e.g.,][]{2004A&A...413..489M,2011ApJ...729....2A,2021MNRAS.502..836S}. It is difficult to obtain useful and valuable information without analyzing quasi-simultaneous multi-waveband data. In this work, we collect quasi-simultaneous multi-waveband data of a sample of $\gamma$-ray quiet FSRQs. By modeling their SEDs with the conventional one-zone leptonic model, we study their physical properties, and explore the possible reason why they are undetected by $Fermi$-LAT. This paper is organized as follows: In Sect.~\ref{sample}, we present the sample collection and $Fermi$-LAT data analysis, the model description is presented in Sect.~\ref{model}. Then, results are shown in Sect.~\ref{results}. Finally, we end with discussions and conclusions in Sect.~\ref{DC}. The cosmological parameters $H_{0}=69.6\ \rm km\ s^{-1}Mpc^{-1}$, $\Omega_{0}=0.29$, and $\Omega_{\Lambda}$= 0.71 \citep{2014ApJ...794..135B} are adopted in this work.

\section{Sample selection and data analysis}\label{sample}
\subsection{Sample selection}
The Candidate Gamma-Ray Blazar Survey (CGRaBS) catalog contains 1625 sources whose radio and X-ray properties are similar to those of the Energetic Gamma Ray Experiment Telescope blazars \citep{1993ApJS...86..629T}, implying these source may be detected by $Fermi$-LAT \citep[][]{2008ApJS..175...97H}. Above all, we find 767 $\gamma$-ray quiet blazars by cross-matching the CGRaBS catalog with the fourth Fermi Large Area Telescope catalog  \citep[4FGL;][]{2020ApJS..247...33A}. Then, we collect all available data and corresponding observation times from the Space Science Data Center (SSDC) SED Builder \citep{2011arXiv1103.0749S}\footnote{http://tools.ssdc.asi.it/SED/}. After that, in order to collect quasi-simultaneous observation data of each source, we search for the intersection of the observation time from infrared to X-ray bands, and ensure that the observation time interval between any two bands does not exceed one month. The radio data available on the SSDC are not taken into account since the most of these data were collected around 1990s, and the conventional one-zone model cannot fit them due to the synchrotron self-absorption. When the one-zone model is used to interpret multiwavelength emission, the emitting region is so compact that the synchrotron emission below the turnover frequency (typically $\nu < 10^{11}$~Hz) is inevitably self-absorbed. Previous studies suggest that the GHz radio data might originate from the extended jet \citep[e.g.,][]{2009MNRAS.399.2041G, 2010MNRAS.402..497G}. This work focuses on fitting the quasi-simultaneous broadband emission from the inner jet; therefore, the radio data below the turnover frequency are not included here.

Finally, 11 $\gamma$-ray quiet FSRQs are selected as our sample with quasi-simultaneous data from infrared to X-ray band. The details of the quasi-simultaneous data for each FSRQ are shown in Table~\ref{tabel1}. No BL Lacs are selected in our sample, since no available quasi-simultaneous data are found based on the above criteria. Meanwhile, we obtain the $Fermi$ GeV upper limits (ULs) of these sources by integrating $\sim$12 years observation (details are given in next subsection).

It should be noted that J2129-1538 in our sample is identified as a $\gamma$-ray loud FSRQ in \citet{2017ApJ...837L...5A}. However, this object has not been included in the recently released 4FGL catalog, and only the ULs are obtained in our GeV data analysis, therefore we still treat this object as a $\gamma$-ray quiet FSRQ in this work.
\subsection{\textsl{Fermi}-LAT data analysis}
$Fermi$-LAT is a pair-production telescope covering the energy band from 20 MeV to >300 GeV\citep{2009ApJ...697.1071A}. We employed the Fermi Science Tool v11r5p3 and an instrument response functions P8R3\_SOURCE\_V2 for this analysis. We used SOURCE class events, converting in both the front and back sections of the LAT. We selected the data collected between 2008 August 4 and 2020 October 26 within a $15^\circ \times 15^\circ$ spatial region of interest (RoI) centered on the position of each target source, with energy range from 100 MeV to 500 GeV. We applied standard quality cuts to our data using $\rm gtmktime$ with expression (DATA\_QUAL>0) \&\& (LAT\_CONFIG==1). To avoid contamination of the $\gamma$-ray photons form the limb of Earth, we discard events with zenith angles $>90^\circ$. The background model includes the sources listed in 4FGL and the isotropic diffuse emission shaped by iso\_P8R3\_SOURCE\_V2\_v01.txt, as well as the Galactic diffuse components described by gll\_iem\_v07.fits. Each target object was fitted to the data following a maximum likelihood approach for binned data and Poisson statistics, which is modeled as a point source with a simple power-law spectrum. Unless otherwise stated, we performed a binned maximum likelihood analysis used a $0.1^\circ \times 0.1^\circ$ pixel size and eight logarithmic energy bins per decade. The significance of each source is quantified with the test statistic (TS),defined as ${\rm TS} = 2(\log L-\log L_0)$, where $L_0$ and $L$ are the maximum likelihood of the background model and of the complete model (i.e., the background model plus the target point source model). No detection ($\rm{TS}>25$) is found for any source in the sample.

The ULs integrating $\sim 12$ years observation were determined by performing a maximum likelihood analysis in 6 logarithmically space energy bins over 0.1-500 GeV.  Within each bin, the spectrum of the target source was modeled as a simple power-law with fixed photon index of 2, and the normalization was allowed to vary, while we set normalization free for diffuse components and for 4FGL sources within $6.5^\circ$ of each target sources. The spectral shape of 4FGL sources were fixed to their parameter given by above broadband analysis. ULs on the flux at $95\%$ confidence level were derived using the Bayesian method in each bin. 

\begin{table*}
\scriptsize 
\caption{The details of the quasi-simultaneous data of our sample.}\label{tabel1}
\centering
\begin{tabular}{ccccccc}
\hline\hline

Source Name & R.A.(J2000) & Decl.(J2000) & $\rm time$ & Telescope & $\nu$ & $\nu F_{\nu}$ \\
 & & & & & (Hz) & $(\rm erg~s^{-1}~cm^{-2})$\\
~(1) & (2) & (3) & (4) & (5) & (6) & (7) \\
\hline
J0106-4034 	&	 01 06 45.12	&	 -40 34 19.9	&	2010.06.16-2010.06.17	&	WISE	&	2.50 	$\times	10^{13}	$	&	$	6.42 	_{	-0.33 	}	^{+	0.33 	}	\times	10^{-13}	$	\\
	&		&		&	2010.06.16-2010.06.17	&	WISE	&	1.36 	$\times	10^{13}	$	&	$	9.74 	_{	-1.24 	}	^{+	1.24 	}	\times	10^{-13}	$	\\
	&		&		&	2010.06.16-2010.06.17	&	WISE	&	8.82 	$\times	10^{13}	$	&	$	4.97 	_{	-0.16 	}	^{+	0.16 	}	\times	10^{-13}	$	\\
	&		&		&	2010.06.16-2010.06.17	&	WISE	&	6.52 	$\times	10^{13}	$	&	$	5.59 	_{	-0.18 	}	^{+	0.18 	}	\times	10^{-13}	$	\\
	&		&		&	2010.06.16-2010.06.17	&	WISE	&	2.50 	$\times	10^{13}	$	&	$	6.21 	_{	-0.29 	}	^{+	0.29 	}	\times	10^{-13}	$	\\
	&		&		&	2010.06.16-2010.06.17	&	WISE	&	1.36 	$\times	10^{13}	$	&	$	9.10 	_{	-1.09 	}	^{+	1.09 	}	\times	10^{-13}	$	\\
	&		&		&	2010.07.03	&	Swift	&	8.56 	$\times	10^{14}	$	&	$	9.78 	_{	-0.80 	}	^{+	0.80 	}	\times	10^{-14}	$	\\
	&		&		&	2010.07.03	&	Swift	&	4.84 	$\times	10^{17}	$	&	$	1.79 	_{	-0.28 	}	^{+	0.28 	}	\times	10^{-13}	$	\\
	&		&		&	2010.07.03	&	Swift	&	4.84 	$\times	10^{17}	$	&	$	1.16 	_{	-0.27 	}	^{+	0.27 	}	\times	10^{-13}	$	\\
	&		&		&	2010.07.03	&	Swift	&	2.42 	$\times	10^{17}	$	&	$	1.67 	_{	-0.28 	}	^{+	0.28 	}	\times	10^{-13}	$	\\
	&		&		&	2010.07.03	&	Swift	&	2.42 	$\times	10^{17}	$	&	$	1.16 	_{	-0.29 	}	^{+	0.29 	}	\times	10^{-13}	$	\\
	&		&		&	2010.07.03	&	Swift	&	1.08 	$\times	10^{18}	$	&	$	3.06 	_{	-0.82 	}	^{+	0.82 	}	\times	10^{-13}	$	\\
	&		&		&	2010.07.03	&	Swift	&	1.33 	$\times	10^{17}	$	&	$	1.48 	_{	-0.28 	}	^{+	0.32 	}	\times	10^{-13}	$	\\
	&		&		&	2010.07.03	&	Swift	&	4.19 	$\times	10^{17}	$	&	$	1.74 	_{	-0.23 	}	^{+	0.23 	}	\times	10^{-13}	$	\\
	&		&		&	2010.07.03	&	Swift	&	3.42 	$\times	10^{17}	$	&	$	1.44 	_{	-0.33 	}	^{+	0.39 	}	\times	10^{-13}	$	\\
	&		&		&	2010.07.03	&	Swift	&	1.08 	$\times	10^{18}	$	&	$	2.37 	_{	-0.53 	}	^{+	0.62 	}	\times	10^{-13}	$	\\
J0140-1532 	&	01 40 04.44	&	 -15 32 55.7	&	2010.07.09-2010.07.10	&	WISE	&	2.50 	$\times	10^{13}	$	&	$	1.11 	_{	-0.05 	}	^{+	0.05 	}	\times	10^{-12}	$	\\
	&		&		&	2010.07.09-2010.07.10	&	WISE	&	1.36 	$\times	10^{13}	$	&	$	1.59 	_{	-0.16 	}	^{+	0.16 	}	\times	10^{-12}	$	\\
	&		&		&	2010.07.23	&	XMM	&	3.75 	$\times	10^{17}	$	&	$	2.06 	_{	-0.35 	}	^{+	0.35 	}	\times	10^{-12}	$	\\
	&		&		&	2010.07.23	&	XMM	&	1.52 	$\times	10^{17}	$	&	$	8.05 	_{	-2.53 	}	^{+	2.53 	}	\times	10^{-13}	$	\\
J0927+3902	&	09 27 03.01 	&	+39 02 20.9	&	2010.04.27-2010.04.28	&	WISE	&	2.50 	$\times	10^{13}	$	&	$	2.81 	_{	-0.08 	}	^{+	0.08 	}	\times	10^{-12}	$	\\
	&		&		&	2010.04.27-2010.04.28	&	WISE	&	1.36 	$\times	10^{13}	$	&	$	3.55 	_{	-0.23 	}	^{+	0.23 	}	\times	10^{-12}	$	\\
	&		&		&	2010.04.27-2010.04.28	&	WISE	&	8.82 	$\times	10^{13}	$	&	$	2.62 	_{	-0.07 	}	^{+	0.07 	}	\times	10^{-12}	$	\\
	&		&		&	2010.04.27-2010.04.28	&	WISE	&	6.52 	$\times	10^{13}	$	&	$	3.53 	_{	-0.09 	}	^{+	0.09 	}	\times	10^{-12}	$	\\
	&		&		&	2010.04.27-2010.04.28	&	WISE	&	2.50 	$\times	10^{13}	$	&	$	2.75 	_{	-0.07 	}	^{+	0.07 	}	\times	10^{-12}	$	\\
	&		&		&	2010.04.27-2010.04.28	&	WISE	&	1.36 	$\times	10^{13}	$	&	$	3.48 	_{	-0.20 	}	^{+	0.20 	}	\times	10^{-12}	$	\\
	&		&		&	2010.04.25	&	Swift	&	4.84 	$\times	10^{17}	$	&	$	1.42 	_{	-0.06 	}	^{+	0.06 	}	\times	10^{-12}	$	\\
	&		&		&	2010.04.25	&	Swift	&	4.84 	$\times	10^{17}	$	&	$	1.16 	_{	-0.12 	}	^{+	0.12 	}	\times	10^{-12}	$	\\
	&		&		&	2010.04.25	&	Swift	&	2.42 	$\times	10^{17}	$	&	$	1.28 	_{	-0.06 	}	^{+	0.06 	}	\times	10^{-12}	$	\\
	&		&		&	2010.04.25	&	Swift	&	2.42 	$\times	10^{17}	$	&	$	1.06 	_{	-0.12 	}	^{+	0.12 	}	\times	10^{-12}	$	\\
	&		&		&	2010.04.25	&	Swift	&	1.08 	$\times	10^{18}	$	&	$	2.04 	_{	-0.18 	}	^{+	0.18 	}	\times	10^{-12}	$	\\
	&		&		&	2010.04.25	&	Swift	&	1.08 	$\times	10^{18}	$	&	$	1.67 	_{	-0.34 	}	^{+	0.34 	}	\times	10^{-12}	$	\\
J0953+3225 	&	09 53 27.95 	&	+32 25 51.5 	&	2010.05.04-2010.05.05	&	WISE	&	2.50 	$\times	10^{13}	$	&	$	1.18 	_{	-0.05 	}	^{+	0.05 	}	\times	10^{-12}	$	\\
	&		&		&	2010.05.04-2010.05.05	&	WISE	&	1.36 	$\times	10^{13}	$	&	$	1.40 	_{	-0.15 	}	^{+	0.15 	}	\times	10^{-12}	$	\\
	&		&		&	2010.05.04-2010.05.05	&	WISE	&	8.82 	$\times	10^{13}	$	&	$	7.48 	_{	-0.22 	}	^{+	0.22 	}	\times	10^{-13}	$	\\
	&		&		&	2010.05.04-2010.05.05	&	WISE	&	6.52 	$\times	10^{13}	$	&	$	1.07 	_{	-0.03 	}	^{+	0.03 	}	\times	10^{-12}	$	\\
	&		&		&	2010.05.04-2010.05.05	&	WISE	&	2.50 	$\times	10^{13}	$	&	$	1.17 	_{	-0.04 	}	^{+	0.04 	}	\times	10^{-12}	$	\\
	&		&		&	2010.05.04-2010.05.05	&	WISE	&	1.36 	$\times	10^{13}	$	&	$	1.34 	_{	-0.13 	}	^{+	0.13 	}	\times	10^{-12}	$	\\
	&		&		&	2010.06.05	&	Swift	&	4.84 	$\times	10^{17}	$	&	$	1.50 	_{	-0.31 	}	^{+	0.31 	}	\times	10^{-13}	$	\\
	&		&		&	2010.06.05	&	Swift	&	2.42 	$\times	10^{17}	$	&	$	8.60 	_{	-1.92 	}	^{+	1.92 	}	\times	10^{-14}	$	\\
	&		&		&	2010.06.05	&	Swift	&	1.08 	$\times	10^{18}	$	&	$	5.93 	_{	-1.61 	}	^{+	1.61 	}	\times	10^{-13}	$	\\
	&		&		&	2010.06.05	&	Swift	&	1.33 	$\times	10^{17}	$	&	$	6.09 	_{	-2.74 	}	^{+	3.85 	}	\times	10^{-14}	$	\\
	&		&		&	2010.06.05	&	Swift	&	4.19 	$\times	10^{17}	$	&	$	1.98 	_{	-0.38 	}	^{+	0.43 	}	\times	10^{-13}	$	\\
	&		&		&	2010.06.05	&	Swift	&	3.42 	$\times	10^{17}	$	&	$	2.45 	_{	-0.71 	}	^{+	0.88 	}	\times	10^{-13}	$	\\
	&		&		&	2010.06.05	&	Swift	&	1.08 	$\times	10^{18}	$	&	$	4.06 	_{	-1.14 	}	^{+	1.40 	}	\times	10^{-13}	$	\\
J1038+0512	&	 10 38 46.78 	&	+05 12 29.1	&	2010.05.25-2010.05.26	&	WISE	&	2.50 	$\times	10^{13}	$	&	$	3.22 	_{	-0.40 	}	^{+	0.40 	}	\times	10^{-13}	$	\\
	&		&		&	2010.05.25-2010.05.26	&	WISE	&	1.36 	$\times	10^{13}	$	&	$	5.59 	_{	-1.29 	}	^{+	1.29 	}	\times	10^{-13}	$	\\
	&		&		&	2010.05.25-2010.05.26	&	WISE	&	8.82 	$\times	10^{13}	$	&	$	3.96 	_{	-0.14 	}	^{+	0.14 	}	\times	10^{-13}	$	\\
	&		&		&	2010.05.25-2010.05.26	&	WISE	&	6.52 	$\times	10^{13}	$	&	$	3.98 	_{	-0.17 	}	^{+	0.17 	}	\times	10^{-13}	$	\\
	&		&		&	2010.05.25-2010.05.26	&	WISE	&	2.50 	$\times	10^{13}	$	&	$	2.78 	_{	-0.37 	}	^{+	0.37 	}	\times	10^{-13}	$	\\
	&		&		&	2010.05.25-2010.05.26	&	WISE	&	1.36 	$\times	10^{13}	$	&	$	5.46 	_{	-1.35 	}	^{+	1.35 	}	\times	10^{-13}	$	\\
	&		&		&	2010.04.28	&	Swift	&	4.84 	$\times	10^{17}	$	&	$	1.15 	_{	-0.21 	}	^{+	0.21 	}	\times	10^{-13}	$	\\
	&		&		&	2010.04.28	&	Swift	&	2.42 	$\times	10^{17}	$	&	$	1.02 	_{	-0.21 	}	^{+	0.21 	}	\times	10^{-13}	$	\\
	&		&		&	2010.04.28	&	Swift	&	1.08 	$\times	10^{18}	$	&	$	2.01 	_{	-0.61 	}	^{+	0.61 	}	\times	10^{-13}	$	\\

\hline
\label{table1}
\end{tabular}
\begin{flushleft}
Notes: Columns from left to right: (1) the source name. (2) right ascension (R.A.). (3) declination (Decl.). (4) the observation date. (5) the observed telescope. (6) the frequency of quasi-simultaneous data. (7) the flux of the corresponding quasi-simultaneous data.
\end{flushleft}
\end{table*}

\addtocounter{table}{-1}

\begin{table*}
\scriptsize 
\caption{-$continued$.}
\centering
\begin{tabular}{ccccccc}
\hline\hline

Source Name & R.A.(J2000) & Decl.(J2000) & $\rm time$ & Telescope & $\nu$ & $\nu F_{\nu}$ \\
 & & & & & (Hz) & $(\rm erg~s^{-1}~cm^{-2})$\\
~(1) & (2) & (3) & (4) & (5) & (6) & (7) \\
\hline

J1423+5055 	&	14 23 14.19 	&	+50 55 37.3	&	2010.06.17-2010.06.21	&	WISE	&	2.50 	$\times	10^{13}	$	&	$	9.56 	_{	-0.19 	}	^{+	0.19 	}	\times	10^{-13}	$	\\
	&		&		&	2010.06.17-2010.06.21	&	WISE	&	1.36 	$\times	10^{13}	$	&	$	2.18 	_{	-0.11 	}	^{+	0.11 	}	\times	10^{-12}	$	\\
	&		&		&	2010.06.17-2010.06.21	&	WISE	&	8.82 	$\times	10^{13}	$	&	$	2.38 	_{	-0.06 	}	^{+	0.06 	}	\times	10^{-12}	$	\\
	&		&		&	2010.06.17-2010.06.21	&	WISE	&	6.52 	$\times	10^{13}	$	&	$	2.50 	_{	-0.06 	}	^{+	0.06 	}	\times	10^{-12}	$	\\
	&		&		&	2010.06.17-2010.06.21	&	WISE	&	2.50 	$\times	10^{13}	$	&	$	1.95 	_{	-0.04 	}	^{+	0.04 	}	\times	10^{-12}	$	\\
	&		&		&	2010.06.17-2010.06.21	&	WISE	&	1.36 	$\times	10^{13}	$	&	$	3.88 	_{	-0.18 	}	^{+	0.18 	}	\times	10^{-12}	$	\\
	&		&		&	2010.07.13	&	Swift	&	4.84 	$\times	10^{17}	$	&	$	1.35 	_{	-0.08 	}	^{+	0.08 	}	\times	10^{-12}	$	\\
	&		&		&	2010.07.13	&	Swift	&	2.42 	$\times	10^{17}	$	&	$	1.26 	_{	-0.07 	}	^{+	0.07 	}	\times	10^{-12}	$	\\
	&		&		&	2010.07.13	&	Swift	&	1.08 	$\times	10^{18}	$	&	$	1.45 	_{	-0.18 	}	^{+	0.18 	}	\times	10^{-12}	$	\\
J2129-1538	&	21 29 12.18	&	 -15 38 41.0	&	2010.05.10-2010.05.11	&	WISE	&	2.50 	$\times	10^{13}	$	&	$	1.66 	_{	-0.06 	}	^{+	0.06 	}	\times	10^{-12}	$	\\
	&		&		&	2010.05.10-2010.05.11	&	WISE	&	1.36 	$\times	10^{13}	$	&	$	2.22 	_{	-0.21 	}	^{+	0.21 	}	\times	10^{-12}	$	\\
	&		&		&	2010.05.10-2010.05.11	&	WISE	&	8.82 	$\times	10^{13}	$	&	$	1.29 	_{	-0.04 	}	^{+	0.04 	}	\times	10^{-12}	$	\\
	&		&		&	2010.05.10-2010.05.11	&	WISE	&	6.52 	$\times	10^{13}	$	&	$	1.00 	_{	-0.03 	}	^{+	0.03 	}	\times	10^{-12}	$	\\
	&		&		&	2010.05.10-2010.05.11	&	WISE	&	2.50 	$\times	10^{13}	$	&	$	1.59 	_{	-0.06 	}	^{+	0.06 	}	\times	10^{-12}	$	\\
	&		&		&	2010.05.10-2010.05.11	&	WISE	&	1.36 	$\times	10^{13}	$	&	$	1.98 	_{	-0.17 	}	^{+	0.17 	}	\times	10^{-12}	$	\\
	&		&		&	2010.05.03	&	Swift	&	4.84 	$\times	10^{17}	$	&	$	1.75 	_{	-0.07 	}	^{+	0.07 	}	\times	10^{-12}	$	\\
	&		&		&	2010.05.03	&	Swift	&	4.84 	$\times	10^{17}	$	&	$	2.07 	_{	-0.15 	}	^{+	0.15 	}	\times	10^{-12}	$	\\
	&		&		&	2010.05.03	&	Swift	&	2.42 	$\times	10^{17}	$	&	$	1.58 	_{	-0.07 	}	^{+	0.07 	}	\times	10^{-12}	$	\\
	&		&		&	2010.05.03	&	Swift	&	2.42 	$\times	10^{17}	$	&	$	1.89 	_{	-0.15 	}	^{+	0.15 	}	\times	10^{-12}	$	\\
	&		&		&	2010.05.03	&	Swift	&	1.08 	$\times	10^{18}	$	&	$	3.74 	_{	-0.23 	}	^{+	0.23 	}	\times	10^{-12}	$	\\
	&		&		&	2010.05.03	&	Swift	&	1.08 	$\times	10^{18}	$	&	$	3.38 	_{	-0.42 	}	^{+	0.42 	}	\times	10^{-12}	$	\\
J2131-1207 	&	21 31 35.26	&	 -12 07 04.8 	&	2010.05.12-2010.05.13	&	WISE	&	2.50 	$\times	10^{13}	$	&	$	3.79 	_{	-0.10 	}	^{+	0.10 	}	\times	10^{-12}	$	\\
	&		&		&	2010.05.12-2010.05.13	&	WISE	&	1.36 	$\times	10^{13}	$	&	$	4.59 	_{	-0.26 	}	^{+	0.26 	}	\times	10^{-12}	$	\\
	&		&		&	2010.05.12-2010.05.13	&	WISE	&	2.50 	$\times	10^{13}	$	&	$	3.67 	_{	-0.10 	}	^{+	0.10 	}	\times	10^{-12}	$	\\
	&		&		&	2010.05.12-2010.05.13	&	WISE	&	1.36 	$\times	10^{13}	$	&	$	4.24 	_{	-0.27 	}	^{+	0.27 	}	\times	10^{-12}	$	\\
	&		&		&	2010.05.09	&	Swift	&	4.84 	$\times	10^{17}	$	&	$	2.35 	_{	-0.15 	}	^{+	0.15 	}	\times	10^{-12}	$	\\
	&		&		&	2010.05.09-2010.05.10	&	Swift	&	4.84 	$\times	10^{17}	$	&	$	2.62 	_{	-0.10 	}	^{+	0.10 	}	\times	10^{-12}	$	\\
	&		&		&	2010.05.09	&	Swift	&	2.42 	$\times	10^{17}	$	&	$	2.10 	_{	-0.14 	}	^{+	0.14 	}	\times	10^{-12}	$	\\
	&		&		&	2010.05.09-2010.05.10	&	Swift	&	2.42 	$\times	10^{17}	$	&	$	2.36 	_{	-0.10 	}	^{+	0.10 	}	\times	10^{-12}	$	\\
	&		&		&	2010.05.09	&	Swift	&	1.08 	$\times	10^{18}	$	&	$	2.69 	_{	-0.33 	}	^{+	0.33 	}	\times	10^{-12}	$	\\
	&		&		&	2010.05.09-2010.05.10	&	Swift	&	1.08 	$\times	10^{18}	$	&	$	3.20 	_{	-0.24 	}	^{+	0.24 	}	\times	10^{-12}	$	\\
J2230-3942	&	22 30 40.28	&	 -39 42 52.1	&	2010.05.14-2010.05.15	&	WISE	&	2.50 	$\times	10^{13}	$	&	$	3.19 	_{	-0.09 	}	^{+	0.09 	}	\times	10^{-12}	$	\\
	&		&		&	2010.05.14-2010.05.15	&	WISE	&	1.36 	$\times	10^{13}	$	&	$	4.84 	_{	-0.25 	}	^{+	0.25 	}	\times	10^{-12}	$	\\
	&		&		&	2010.05.14-2010.05.15	&	WISE	&	8.82 	$\times	10^{13}	$	&	$	2.43 	_{	-0.07 	}	^{+	0.07 	}	\times	10^{-12}	$	\\
	&		&		&	2010.05.14-2010.05.15	&	WISE	&	6.52 	$\times	10^{13}	$	&	$	2.85 	_{	-0.07 	}	^{+	0.07 	}	\times	10^{-12}	$	\\
	&		&		&	2010.05.14-2010.05.15	&	WISE	&	2.50 	$\times	10^{13}	$	&	$	3.17 	_{	-0.08 	}	^{+	0.08 	}	\times	10^{-12}	$	\\
	&		&		&	2010.05.14-2010.05.15	&	WISE	&	1.36 	$\times	10^{13}	$	&	$	4.78 	_{	-0.24 	}	^{+	0.24 	}	\times	10^{-12}	$	\\
	&		&		&	2010.05.09	&	Swift	&	4.84 	$\times	10^{17}	$	&	$	2.49 	_{	-0.10 	}	^{+	0.10 	}	\times	10^{-12}	$	\\
	&		&		&	2010.05.09	&	Swift	&	4.84 	$\times	10^{17}	$	&	$	2.90 	_{	-0.16 	}	^{+	0.16 	}	\times	10^{-12}	$	\\
	&		&		&	2010.05.09	&	Swift	&	2.42 	$\times	10^{17}	$	&	$	2.22 	_{	-0.10 	}	^{+	0.10 	}	\times	10^{-12}	$	\\
	&		&		&	2010.05.09	&	Swift	&	2.42 	$\times	10^{17}	$	&	$	2.61 	_{	-0.15 	}	^{+	0.15 	}	\times	10^{-12}	$	\\
	&		&		&	2010.05.09	&	Swift	&	1.08 	$\times	10^{18}	$	&	$	3.02 	_{	-0.25 	}	^{+	0.25 	}	\times	10^{-12}	$	\\
	&		&		&	2010.05.09	&	Swift	&	1.08 	$\times	10^{18}	$	&	$	3.11 	_{	-0.36 	}	^{+	0.36 	}	\times	10^{-12}	$	\\
J2246-1206 	&	22 46 18.23	&	 -12 06 51.3	&	2010.05.28-2010.05.29	&	WISE	&	2.50 	$\times	10^{13}	$	&	$	1.37 	_{	-0.06 	}	^{+	0.06 	}	\times	10^{-12}	$	\\
	&		&		&	2010.05.28-2010.05.29	&	WISE	&	1.36 	$\times	10^{13}	$	&	$	1.65 	_{	-0.20 	}	^{+	0.20 	}	\times	10^{-12}	$	\\
	&		&		&	2010.05.28-2010.05.29	&	WISE	&	8.82 	$\times	10^{13}	$	&	$	1.56 	_{	-0.04 	}	^{+	0.04 	}	\times	10^{-12}	$	\\
	&		&		&	2010.05.28-2010.05.29	&	WISE	&	6.52 	$\times	10^{13}	$	&	$	1.95 	_{	-0.05 	}	^{+	0.05 	}	\times	10^{-12}	$	\\
	&		&		&	2010.05.28-2010.05.29	&	WISE	&	2.50 	$\times	10^{13}	$	&	$	1.35 	_{	-0.05 	}	^{+	0.05 	}	\times	10^{-12}	$	\\
	&		&		&	2010.05.28-2010.05.29	&	WISE	&	1.36 	$\times	10^{13}	$	&	$	1.54 	_{	-0.17 	}	^{+	0.17 	}	\times	10^{-12}	$	\\
	&		&		&	2010.05.10	&	Swift	&	4.84 	$\times	10^{17}	$	&	$	6.42 	_{	-0.93 	}	^{+	0.93 	}	\times	10^{-13}	$	\\
	&		&		&	2010.05.10	&	Swift	&	2.42 	$\times	10^{17}	$	&	$	5.90 	_{	-0.89 	}	^{+	0.89 	}	\times	10^{-13}	$	\\
	&		&		&	2010.05.10	&	Swift	&	1.08 	$\times	10^{18}	$	&	$	8.81 	_{	-2.33 	}	^{+	2.33 	}	\times	10^{-13}	$	\\
J2251-3827  	&	22 51 19.03 	&	-38 27 07.2	&	2010.05.19-2010.05.20	&	WISE	&	2.50 	$\times	10^{13}	$	&	$	1.18 	_{	-0.05 	}	^{+	0.05 	}	\times	10^{-12}	$	\\
	&		&		&	2010.05.19-2010.05.20	&	WISE	&	1.36 	$\times	10^{13}	$	&	$	1.85 	_{	-0.19 	}	^{+	0.19 	}	\times	10^{-12}	$	\\
	&		&		&	2010.05.19-2010.05.20	&	WISE	&	8.82 	$\times	10^{13}	$	&	$	1.59 	_{	-0.04 	}	^{+	0.04 	}	\times	10^{-12}	$	\\
	&		&		&	2010.05.19-2010.05.20	&	WISE	&	6.52 	$\times	10^{13}	$	&	$	1.47 	_{	-0.04 	}	^{+	0.04 	}	\times	10^{-12}	$	\\
	&		&		&	2010.05.19-2010.05.20	&	WISE	&	2.50 	$\times	10^{13}	$	&	$	1.17 	_{	-0.04 	}	^{+	0.04 	}	\times	10^{-12}	$	\\
	&		&		&	2010.05.19-2010.05.20	&	WISE	&	1.36 	$\times	10^{13}	$	&	$	1.73 	_{	-0.14 	}	^{+	0.14 	}	\times	10^{-12}	$	\\
	&		&		&	2010.05.21	&	XMM	&	3.75 	$\times	10^{17}	$	&	$	4.20 	_{	-1.71 	}	^{+	1.71 	}	\times	10^{-13}	$	\\
	&		&		&	2010.05.21	&	XMM	&	1.52 	$\times	10^{17}	$	&	$	4.02 	_{	-1.27 	}	^{+	1.27 	}	\times	10^{-13}	$	\\

\hline
\end{tabular}
\\
\end{table*}

\section{Model Description}\label{model}
In this section, all parameters are measured in the comoving frame, unless specified otherwise. We fit the SEDs of our sample with the conventional one-zone EC model. In this model, the dissipation region is considered as a single spherical region composed of a plasma of charged particles in a uniformly magnetic field $B$ with radius $R$ and moving with bulk Lorentz factor $\Gamma=(1-\beta_{\Gamma}^2)^{-1/2}$, where $\beta_{\Gamma}c$ is the jet speed, along the jet, at a viewing angle $\theta_{\rm obs}$ with respect to observers' line of sight. Due to the beaming effect, the observed radiation is strongly boosted by a relativistic Doppler factor $\delta=[\Gamma(1-\beta_{\Gamma} \rm cos\theta_{\rm obs})]^{-1}$. In this work, by assuming $\theta_{\rm obs} \lesssim 1/\Gamma$ for our sample, we have $\delta \approx \Gamma$. In modeling, accelerated relativistic electrons are assumed to be injected into the dissipation region with a broken power-law distribution at a constant rate \citep{2010MNRAS.402..497G}, i.e.,
\begin{equation}\label{Q}
   Q(\gamma)=Q_{\rm 0} \gamma^{-p_{\rm 1}} [1+(\frac{\gamma}{\gamma_{\rm b}})^{p_{\rm 2}-p_{\rm 1}}]^{-1}, \   \gamma_{\rm min}<\gamma_{\rm b}<\gamma_{\rm max},
\end{equation}
where $\gamma$ is the electron Lorentz factor, $\gamma_{\rm min}$, $\gamma_{\rm b}$ and $\gamma_{\rm max}$ are the minimum, break and maximum electron Lorentz factors, respectively, $p_{\rm 1}$ and $p_{\rm 2}$ are the spectral indices before and after $\gamma_{\rm b}$, and $Q_{\rm0}$ is a normalization constant in units of $\rm s^{-1}~cm^{-3}$. We can calculate $Q_{\rm 0}$ from 
\begin{equation}\label{inj}
\int{Q(\gamma) \gamma m_{\rm e} c^2 d\gamma} = \frac{3L_{\rm e, inj}}{4{\rm \pi} R^3},
\end{equation} 
where $L_{\rm e,inj}$ is the electron injection luminosity, $m_{\rm e}$ is the electron rest mass, and $c$ is the speed of light. When the injection of electrons is balanced with radiative cooling and/or particle escape, a steady-state electron energy distribution (EED) is achieved. It can be written as
\begin{equation}
   N(\gamma)\approx Q(\gamma)t_{\rm e},
   \label{eq:quadratic2}
\end{equation}
where $t_{\rm e}$ = min\{$t_{\rm cool},t_{\rm dyn}$\}. $t_{\rm dyn}$ = $R/c$ is the dynamical timescale and $t_{\rm cool}$ = 3$m_{\rm e} c/[4(U_{\rm B}+\kappa_{\rm KN}U_{\rm ph})\sigma_{\rm T}\gamma]$ is the radiative cooling timescale, where $\sigma_{\rm T}$ is the Thomson scattering cross-section, $\kappa_{\rm KN}$ is a numerical factor accounting for Klein-Nishina effects \citep[]{2010NJPh...12c3044S}, $U_{\rm B}$ = $B^2/(8{\pi})$ is the energy density of the magnetic field, and $U_{\rm ph}=U_{\rm syn}+U_{\rm ext}$ is the energy density of the soft photons. $U_{\rm syn}$ is the energy density of the synchrotron photons, and $U_{\rm ext}$ is the energy density of external photons. In the environment of blazar jet, external photons are mainly from the BLR and DT. The energy density of the BLR ($U_{\rm BLR}$) and the DT ($U_{\rm DT}$) in the jet comoving frame can be estimated as \citep[]{2012ApJ...754..114H}
\begin{equation}\label{BLR}
   U_{\rm BLR}=\frac{\eta_{\rm BLR}\Gamma^2L_{\rm d}}{4\pi r_{\rm BLR}^2 c [1+(r/r_{\rm BLR})^3]}
\end{equation}
and
\begin{equation}\label{DT}
   U_{\rm DT}=\frac{\eta_{\rm DT}\Gamma^2 L_{\rm d}}{4\pi r_{\rm DT}^2 c [1+(r/r_{\rm DT})^4]},
\end{equation}
where $r$ is the distance of the dissipation region from the central super-massive black hole (SMBH) in the AGN frame, $\eta_{\rm BLR}$ = $\eta_{\rm DT}$ = 0.1 are the fractions of the disk luminosity $L_{\rm d}$ reprocessed into the BLR and DT radiation, respectively, $r_{\rm BLR}$ = 0.1($L_{\rm d}/10^{46}$ erg s$^{-1})^{1/2}$ pc and $r_{\rm DT}$ = 2.5($L_{\rm d}/10^{46}$ erg s$^{-1})^{1/2}$ pc are the characteristic distances of the BLR and DT, respectively. After obtaining the steady state EED $N(\gamma)$, we adopt the public Python package \texttt{NAIMA}\footnote{https://naima.readthedocs.io/en/latest/} to calculate the synchrotron, SSC and EC emissions from jets and correct the GeV-TeV spectrum absorbed by the extragalactic background light \citep[]{2015ICRC...34..922Z}.

The model contains 11 free parameters: $B$, $R$, $p_{\rm 1}$, $p_{\rm 2}$, $\gamma_{\rm min}$, $\gamma_{\rm b}$, $\gamma_{\rm max}$, $L_{\rm e,inj}$, $\delta$, $r$ and $L_{\rm d}$. Here we set $\gamma_{\rm min}$ = 48 \citep[]{2014ApJ...788..104Z}, and $\gamma_{\rm max}$ = 2$\times10^6$ \citep[]{2005A&A...432..401G}, since they have little affection for our fitting results. In addition, we estimate $L_{\rm d}$ to be twice the peak luminosity of the thermal component in the optical-UV band \citep[]{2015MNRAS.448.1060G}, which can be obtained from the archive data on the SSDC.

In the modeling, we adopt the Markov chain Monte Carlo (MCMC) method to obtain the best-fit parameters. We apply the \texttt{emcee} Python package\footnote{https://emcee.readthedocs.io/en/stable/} \citep[version 3.0.2,][]{2013PASP..125..306F}, which implements the affine-invariant ensemble sampler of \cite{2010CAMCS...5...65G}, to perform the MCMC algorithm. In the application of the MCMC method, in addition to fitting the quasi-simultaneous data from infrared to X-ray, the $Fermi$ GeV ULs are also regarded as data points and fitted as much as possible, while ensuring that the model predicted flux will not overshoot any GeV ULs. This helps constrain the parameters and show more clearly how the parameters affect the $\gamma$-ray emission in Sect.~\ref{results}. However, since most of the free parameters are coupled to each other, it is difficult to get a unique best fitting parameter set with MCMC method \citep[e.g.,][]{2020PASJ...72...42Y}. Therefore, some parameters need to be further fixed. When applying the one-zone EC model to fit the SEDs of FSRQs, it is generally accepted that the low energy component is from synchrotron emission, X-ray data are ascribed to the SSC emission, and GeV data are explained by the EC emission \citep[e.g.,][]{2020ApJS..248...27T, 2021MNRAS.503.2523H}. Therefore, under the premise that the flux of each component can be well limited by the quasi-simultaneous data, relations between the fluxes from different components in the Thomson regime can be obtained as
\begin{equation}
  \frac{\nu F_{\rm\nu, X-ray}}{\nu F_{\rm\nu, optical}}\approx \frac{\nu F_{\rm\nu, SSC}}{\nu F_{\rm\nu, syn}}=\frac{U_{\rm syn}}{U_{\rm B}},
   \label{eq:quadratic5}
\end{equation}
in which an anticorrelation can be found, i.e., $B$ $\propto$ $R^{-1} \delta^{-2}$, and 
\begin{equation}
  \frac{\nu F_{\rm\nu, GeV}}{\nu F_{\rm\nu, optical}}\approx \frac{\nu F_{\rm\nu, EC}}{\nu F_{\rm\nu, syn}}=\frac{U_{\rm ext}}{U_{\rm B}},
   \label{eq:quadratic6}
\end{equation}
in which we find $B$ $\propto$ $\delta r^{n/2}$, where the value of $n$ depends on the specific location $r$ of the dissipation region, as shown in Eq.~\ref{BLR}, \ref{DT}. Similar relations also can be obtained in the Klein-Nishina regime. Therefore, it can be seen that if we fix $\delta$ and $r$, or $\delta$ and $R$, we can derive $B$ and the remaining parameter.

In this work, we focus on studying why $\gamma$-ray quiet FSRQs are not detected by $Fermi$-LAT. Theoretically, there are three parameters ($\delta$, $r$, $L_{\rm e, inj}$) that would have enormous impacts on observated $\gamma$-ray flux (details can be found in the next section). Among them, $L_{\rm e, inj}$ can be constrained well in the fitting. However, according to the above discussion, the other two parameters are degenerated with the magnetic field. In order to reduce the effects of degeneracy and further study in detail how these parameters affect the $\gamma$-ray emission, we fix $\delta$ and $r$ in the modeling with MCMC method. Here we set $\delta=10$ which is a typical value suggested in observations \citep{2009A&A...494..527H}, and since a lot of studies \citep[e.g.,][]{2018MNRAS.477.4749C} show that most of the dissipation region of FSRQ are located outside the BLR and within the DT, we take log$r$ = [log$(r_{\rm BLR})$+log$(r_{\rm DT})$]/2. Note that this location of the dissipation region is suggested from modeling the SED of $\gamma$-ray loud FSRQs, and may not be appropriate for the $\gamma$-ray quiet FSRQs because of the different observed features.

\section{Results}\label{results}
The one-zone EC leptonic model is applied to fit the SEDs of 11 $\gamma$-ray quiet FSRQs. The model predicted SEDs for the parameters that obtained by the MCMC method shown in Table~\ref{table2} are presented in Figure~\ref{fig1}. GeV ULs are the threshold of whether $\gamma$-ray can be detected for each source. During the fitting, the $Fermi$ GeV ULs are regarded as data points and are also fitted, which represent the maximum flux that keeps $\gamma$-ray from being detected. 

In the one-zone leptonic model, the observed radiative luminosity can be calculated as
\begin{equation}\label{L}
   L_{\rm jet}^{\rm obs}=\frac{4}{3} \pi R^3 \delta^4 \int f_{\rm e} \gamma m_{\rm e} c^2 Q(\gamma) d \gamma,
\end{equation}
where $f_{\rm e}$ = min\{$\frac{t_{\rm dyn}}{t_{\rm cool}},1$\} is the radiative efficiency of relativistic electrons. Substituting Eq.~\ref{Q},~\ref{inj} into Eq.~\ref{L}, we have
\begin{equation}
   L_{\rm jet}^{\rm obs}=L_{\rm e,inj} \delta^4 \int f_{\rm e} \frac{\gamma^{1-p_1}[1+(\frac{\gamma}{\gamma_{\rm b}})^{ p_2-p_1}]^{-1}}{\int \gamma^{1-p_1}[1+(\frac{\gamma}{\gamma_{\rm b}})^{p_2-p_1}]^{-1} d\gamma} d \gamma,
   \label{eq:quadratic9}
\end{equation}
in which $U_{\rm B}$, $U_{\rm syn}$ and $U_{\rm ext}$ in $t_{\rm cool}$ affect the flux of synchrotron, SSC and EC radiation, respectively. From Eq.~\ref{eq:quadratic9}, it can be seen that $\delta$, $L_{\rm e,inj}$ and $r$ have enormous direct impacts on emitted $\gamma$-ray flux, while the influence of the rest free parameters on emitted $\gamma$-ray flux is relatively minor and indirect. $\gamma_{\rm b}$, $p_1$ and $p_2$ are free parameters related to EED. They can directly affect the electron energy distribution and have a certain influence on the photon energy distribution. For example, increasing $p_2$ will steepen the IC spectrum in the $\gamma$-ray band, making it undetectable by $Fermi$-LAT \citep{2017ApJ...851...33P}. However, under the constraints of the quasi-simultaneous data, they are well restricted and have relatively minor effect on the SED. According to $f_{\rm e}$, the value of $R$ will affect the cooling efficiency of the relativistic electrons. It can be seen from Eq.~\ref{eq:quadratic9} that when the value of $R$ makes $t_{\rm dyn}$ greater than $t_{\rm cool}$, the radiation generated by the relativistic electrons is not related to $R$. At the same time, since $\gamma$-ray radiation is mainly caused by the high-energy electrons, their cooling timescale ($t_{\rm cool} \propto \gamma ^{-1}$) is usually shorter than the dynamical timescale, unless considering a very compact radiation region. Therefore adjusting $R$ has little the influence on the $\gamma$-ray flux. The magnetic field $B$ is related to the energy density of the magnetic field $U_{\rm B}$, so adjusting $B$ will directly affect the flux of synchrotron radiation. Meanwhile, adjusting $B$ will change the ratio of $U_{\rm B}$ to $U_{\rm ext}$, therefore it will also have a certain impact on the $\gamma$-ray flux. However, since a significant difference between the synchrotron flux of $\gamma$-ray loud FSRQs and $\gamma$-ray quiet FSRQs is not found (more on that later), $B$ is also not regarded as the possible reason why the $\gamma$-ray quiet FSRQs have no $\gamma$-ray detection. In the following, we will adjust $\delta$, $L_{\rm e,inj}$ and $r$ that have direct impacts on the $\gamma$-ray flux based on the results that obtained from the MCMC method (the black solid line in Figure~\ref{fig1}) and show how they affect the $\gamma$-ray flux.

\begin{table*}
\caption{The best-fit model parameters given by MCMC method.}
\centering
\begin{tabular}{cccccccccc}
\hline\hline
Source Name & $z$ & log$L_{\rm d}$ & $r$ & $B$ & log$R$ & $p_{1}$ & $p_{2}$ & log$\gamma_{\rm b}$ & log$L_{\rm {e,inj}}$ \\
 & & $(\rm erg~s^{-1})$ & $(\rm pc)$ & $(\rm G)$ & $(\rm cm)$ & & & & $(\rm erg~s^{-1})$ \\
~(1) & (2) & (3) & (4) & (5) & (6) & (7) & (8) & (9) & (10) \\
\hline\\
J0106-4034	&	0.584	&	45.62	&	0.323	&	$1.209_{-0.011}^{+0.010}$	&	$16.299_{-0.008}^{+0.008}$	&	$1.543_{-0.033}^{+0.059}$	&	$3.729_{-0.038}^{+0.040}$	&	$2.619_{-0.018}^{+0.021}$	&	$42.596_{ -0.006}^{+0.007}$	\\
J0140-1532	&	0.819	&	46.73	&	1.155	&	$1.276_{-0.017}^{+0.039}$	&	$15.853_{-0.027}^{+0.013}$	&	$0.646_{-0.050}^{+0.054}$	&	$5.889_{-0.283}^{+0.088}$	&	$2.811_{-0.012}^{+0.009}$	&	$43.031_{ -0.007}^{+0.003}$	\\
J0927+3902	&	0.695	&	46.51	&	0.902	&	$1.709_{-0.007}^{+0.027}$	&	$16.644_{-0.027}^{+0.014}$	&	$1.509_{-0.007}^{+0.058}$	&	$3.990_{-0.051}^{+0.007}$	&	$2.620_{-0.008}^{+0.014}$	&	$43.282_{ -0.010}^{+0.009}$	\\
J0953+3225	&	1.574	&	47.27	&	2.153	&	$1.745_{-0.054}^{+0.076}$	&	$16.230_{-0.025}^{+0.019}$	&	$0.276_{-0.054}^{+0.080}$	&	$3.664_{-0.052}^{+0.063}$	&	$2.953_{-0.014}^{+0.017}$	&	$43.159_{ -0.006}^{+0.007}$	\\
J1038+0512	&	0.473	&	45.42	&	0.256	&	$0.610_{-0.008}^{+0.009}$	&	$16.200_{-0.046}^{+0.062}$	&	$1.867_{-0.166}^{+0.095}$	&	$3.037_{-0.027}^{+0.073}$	&	$2.577_{-0.101}^{+0.096}$	&	$42.565_{ -0.022}^{+0.035}$	\\
J1423+5055	&	0.276	&	45.84	&	0.418	&	$2.782_{-0.100}^{+0.088}$	&	$15.388_{-0.019}^{+0.021}$	&	$2.336_{-0.066}^{+0.082}$	&	$3.282_{-0.055}^{+0.129}$	&	$2.591_{-0.103}^{+0.189}$	&	$42.358_{ -0.012}^{+0.015}$	\\
J2129-1538	&	3.280	&	48.02	&	5.137	&	$1.197_{-0.035}^{+0.057}$	&	$16.704_{-0.029}^{+0.025}$	&	$1.550_{-0.033}^{+0.055}$	&	$3.891_{-0.106}^{+0.075}$	&	$3.273_{-0.027}^{+0.032}$	&	$44.269_{ -0.015}^{+0.013}$	\\
J2131-1207	&	0.501	&	46.92	&	1.446	&	$1.607_{-0.067}^{+0.057}$	&	$16.110_{-0.016}^{+0.019}$	&	$2.212_{-0.110}^{+0.068}$	&	$3.497_{-0.102}^{+0.144}$	&	$2.999_{-0.144}^{+0.122}$	&	$43.043_{ -0.011}^{+0.010}$	\\
J2230-3942	&	0.318	&	45.68	&	0.347	&	$1.923_{-0.040}^{+0.038}$	&	$15.770_{-0.025}^{+0.034}$	&	$1.961_{-0.113}^{+0.166}$	&	$3.594_{-0.076}^{+0.069}$	&	$2.628_{-0.084}^{+0.073}$	&	$42.694_{ -0.019}^{+0.023}$	\\
J2246-1206	&	0.630	&	46.5	&	0.894	&	$0.958_{-0.009}^{+0.013}$	&	$16.386_{-0.012}^{+0.034}$	&	$1.661_{-0.046}^{+0.049}$	&	$3.007_{-0.005}^{+0.021}$	&	$2.588_{-0.024}^{+0.009}$	&	$43.022_{ -0.009}^{+0.020}$	\\
J2251-3827	&	0.135	&	45.58	&	0.309	&	$1.985_{-0.044}^{+0.071}$	&	$15.300_{-0.015}^{+0.023}$	&	$2.175_{-0.088}^{+0.147}$	&	$3.247_{-0.076}^{+0.067}$	&	$2.629_{-0.155}^{+0.135}$	&	$41.802_{ -0.013}^{+0.025}$	\\

\hline
\label{table2}
\end{tabular}
\\
\end{table*}

Adjusting $\delta$, $L_{\rm e,inj}$ and $r$ will have different impacts on the synchrotron, SSC and EC emission. Due to the beaming effect, the full waveband radiation will be amplified by $\delta^4$. In addition, since the energy of photons radiated in the leptonic model come from the relativistic electrons, adjusting $L_{\rm e,inj}$ will also have a great impact on the full waveband radiation. For $r$, it is only related to the energy densities of external photon fields $U_{\rm ext}$. Therefore, $r$ has enormous impact on the EC flux, but has minor impact on the synchrotron and SSC flux. Unless $U_{\rm ext}$ changes from being much smaller than $U_{\rm B}$ to being much higher than $U_{\rm B}$, it will cause the emission that should be radiated through synchrotron process to be radiated through the EC process, decreasing flux of synchrotron component \citep{2019ApJ...886...23X}. However, this would not happen to $\gamma$-ray quiet FSRQs, since no $\gamma$-ray is detected originally, which means that $U_{\rm ext}$ is lower than $U_{\rm B}$ (as shown by the red dashed lines in Figure~\ref{fig1}). In addition, decreasing $U_{\rm ext}$ will reduce the EC flux at the GeV band, but does not increase the synchrotron flux significantly, just as shown by the red dotted line in Figure~\ref{fig1}.


Taking the parameters in Table~\ref{table2} as the standard, we adjust $\delta$, $L_{\rm e,inj}$ and $r$ to check how they affect synchrotron, SSC and EC radiation. The model predicted results are also presented in Figure~\ref{fig1} with colored dashed and dotted lines. It can be seen that, as we analyzed above, adjusting $\delta$ and $L_{\rm e,inj}$ will cause the flux of synchrotron, SSC and EC radiation to rise and fall together (see the green/purple dashed and dotted lines in Figure~\ref{fig1}), while adjusting $r$ will only affect the EC radiation significantly (see the red dashed and dotted lines in Figure~\ref{fig1}). Therefore, as suggested by previous studies \citep{2014A&A...562A..64W, 2015ApJ...810L...9L, 2017ApJ...851...33P}, if having smaller $\delta$ or $L_{\rm e, inj}$ is the main reason that $\gamma$-ray quiet FSRQs do not detect $\gamma$-ray, the synchrotron flux of these $\gamma$-ray quiet FSRQs would also be much lower than that of $\gamma$-ray loud FSRQs. Here we select 98 $Fermi$-2LAC and 60 $Fermi$-4LAC FSRQs with quasi-simultaneous SEDs in \cite{2016MNRAS.463.3038X} and \cite{2020ApJS..248...27T}, respectivly, as control samples to compare the synchrotron peak flux between our $\gamma$-ray quiet FSRQs and their $\gamma$-ray loud FSRQs (see Figure~\ref{fig2}). A Kolmogorov-Smirnov test indicates that the synchrotron peak flux distribution in our sample is not significantly different from that in \cite{2016MNRAS.463.3038X} and \cite{2020ApJS..248...27T} ($p$ = 0.770 and 0.369 respectively).
Therefore, our results suggest that $r$ is more likely to be the main difference between the $\gamma$-ray quiet FSRQs and the $\gamma$-ray loud FSRQs. Meanwhile, as analyzed above and shown in Figure~\ref{fig1}, 
increasing $r$ will make the model predicted GeV flux lower than $Fermi$ GeV ULs, therefore it further suggests that the locations of the dissipation region of the $\gamma$-ray quiet FSRQs are farther away from SMBH than that of $\gamma$-ray loud FSRQs. A lot of previous studies suggest that the dissipation regions of $\gamma$-ray loud blazars are located near the parsec-scale DT \citep[e.g.,][]{2020NatCo..11.5475H, 2020ApJS..248...27T}. Therefore, we believe that the dissipation of the $\gamma$-ray quiet FSRQs occurs far beyond the DT, so that the energy densities of external photon fields are quite low, and the EC radiation that mainly contributes to the GeV flux is too weak to be detected.

\section{Discussions and Conclusions}\label{DC}
To summarize, we study the staple reason for the non-$\gamma$-ray detection of a sample of $\gamma$-ray quiet FSRQs. We find that the synchrotron flux of $\gamma$-ray quiet FSRQs is not significantly different from that of $\gamma$-ray loud FSRQs. It suggests that the difference in the $\gamma$-ray band is more likely caused by the location of the dissipation region. In this work, we focus on which physical parameter has a decisive influence, therefore we boldly fixed the rest parameters. Of course, it is undeniable that a joint influence of multiple parameters is also a possible explanation. Previous studies \citep[e.g.,][]{2014A&A...562A..64W, 2015ApJ...810L...9L, 2017ApJ...851...33P} suggests that having a smaller Doppler factor is the most likely reason why a great many blazars are not detected with $\gamma$-ray. However, if the electron injection luminosity and magnetic field are not significantly enhanced simultaneously, the synchrotron flux will be significantly reduced, as shown by the purple dotted line in Figure~\ref{fig1}. Therefore, the non-$\gamma$-ray detection of $\gamma$-ray quiet FSRQs is unlikely to be caused by the Doppler factor alone, at least it also needs to introduce larger electron injection luminosity or stronger magnetic field while ensuring that their synchrotron flux is still comparable to that of the $\gamma$-ray loud FSRQs.

On the other hand, $\gamma$-ray emitted in the dissipation region might be absorbed due to the internal $\gamma\gamma$ absorption. For $Fermi$ satellite, its detection energy range is from 50~MeV to 1~TeV \citep{2020ApJS..247...33A}, therefore the $\gamma\gamma$ absorption optical depth at 50 MeV $\tau_{\gamma \gamma}^{50~\rm MeV}$ should be $\geqslant 1$. For each FSRQ in our sample, the soft photon's energies in observer's frame that annihilate $\gamma$-ray at 50~MeV are estimated with \citep{2008ApJ...686..181F} 
\begin{equation}
\overline E_{\rm obs}=\frac{2 (m_{\rm e} c^2)^2}{(1+z)^2 50~\rm MeV} , 
\end{equation}
and shown in Table~\ref{table3}. It can be seen that the obtained soft photon's energy for each FSRQ is about a few keV. For FSRQs, the number density of the soft photons at the keV band from the dissipation region inside is very low and cannot absorb $\gamma$-ray. Therefore, if $\gamma$-ray is indeed absorbed, the soft photons must come from the external photon field. In AGN environment, the hot corona surrounding the accretion disk could emit X-ray photons from 0.1 to 100~keV \citep{2018ApJ...866..124K,2018MNRAS.480.1819R}. If the X-ray photons emitted by hot corona are boosted in the dissipation region, the dissipation region need to form in the corona region. Therefore, we assume that the dissipation region is near the jet base with a distance $r=5\times 10^{14}~\rm cm$ that comparable to a few times larger than the Schwarzschild radius of the central SMBH. If $\tau_{\gamma \gamma}^{50~\rm MeV}=1$, the required minimum flux of soft photons to absorb $\gamma$-ray can be given by \citep{2021ApJ...906...51X}
\begin{equation}
\nu F_{\nu,\rm soft}=\frac{r c \overline E_{\rm obs}\tau_{\gamma \gamma}^{50~\rm MeV}}{\sigma_{\gamma\gamma} D_{\rm L}^2},
   \label{eq10}
\end{equation}
where $\sigma_{\gamma\gamma} \approx 1.68 \times 10^{-25}~\rm cm^2$ is the cross section for $\gamma\gamma$ annihilation, $\nu F_{\nu,\rm soft}$ is the flux of soft photons in the observer's frame and $D_{\rm L}$ is the luminosity distance. The calculated minimum flux $\nu F_{\nu,\rm soft}$ is shown in Table~\ref{table3}, and it can be seen that $\nu F_{\nu,\rm soft}$ for each FSRQ is lower than the flux of observational data points in the X-ray band, which is consistent with the fact that the X-ray emission from the hot corona is not directly observed. Therefore, if the jet's dissipation can occurs in the hot corona region, no $\gamma$-ray can be observed as well. On the other hand, \cite{2021ApJ...906...51X} indicates that high-energy neutrino flares can be produced in such a region where the jet's dissipation occurs in the hot corona, although the occurrence probability is quite low. At present, there are still a large number of jet-dominated AGNs have not been detected in the $\gamma$-ray band. If a considerable part of their jet dissipation occurs in the X-ray corona, then the jet-dominated $\gamma$-ray quiet AGNs may also contribute to the high-energy neutrino background.


\begin{table}
\scriptsize 
\caption{The maximum energy of the soft photons and the corresponding observation flux for $\gamma\gamma$ absorption.}
\tabcolsep 5mm
\begin{tabular}{cccc}
\hline\hline
Source Name & $z$ & $\overline E_{\rm obs}$ & $\nu F_{\nu,\rm soft}$  \\
& & (keV) & $(\rm erg~s^{-1}~cm^{-2})$ \\
~(1) & (2) & (3) & (4)\\
\hline
J0106-4034 & 0.584 & 4.164 & 5.62$\times 10^{-15}$\\
J0140-1532 & 0.819 & 3.158 & 1.86$\times 10^{-15}$\\
J0927+3902 & 0.695 & 3.636 & 3.21$\times 10^{-15}$\\
J0953+3225 & 1.574 & 1.577 & 1.85$\times 10^{-16}$\\
J1038+0512 & 0.473 & 4.815 & 1.08$\times 10^{-14}$\\
J1423+5055 & 0.276 & 6.417 & 5.08$\times 10^{-14}$\\
J2129-1538 & 3.280 & 0.570 & 1.12$\times 10^{-17}$\\
J2131-1207 & 0.501 & 4.637 & 9.06$\times 10^{-15}$\\
J2230-3942 & 0.318 & 6.014 & 3.43$\times 10^{-14}$\\
J2246-1206 & 0.630 & 3.932 & 4.42$\times 10^{-15}$\\
J2251-3827 & 0.135 & 8.110 & 3.16$\times 10^{-13}$\\

\hline
\end{tabular}
\label{table3}
\\
\end{table}

\section{Acknowledgments}
We thank the anonymous referee for insightful comments and constructive suggestions. Part of this work is based on archival data, software, or online services provided by the SPACE SCIENCE DATA CENTER (SSDC). This work is supported by the Ph.D. Research start-up fund of Zhejiang Normal University (Grant No. YS304320082). Z.H.X. acknowledges support from the Joint Research Fund in Astronomy (grant Nos. 10978019, U1431123) under cooperative agreement between the National Natural Science Foundation of China (NSFC) and Chinese Academy of Sciences (CAS). L.M.D. is supported by the Provincial Natural Science Foundation of Yunnan (grant No. 2019FB009).

\section*{data availability}
The data underlying this article will be shared on reasonable request to the corresponding author.


\onecolumn
\begin{figure}
\centering
\includegraphics[width=7cm,height=4cm]{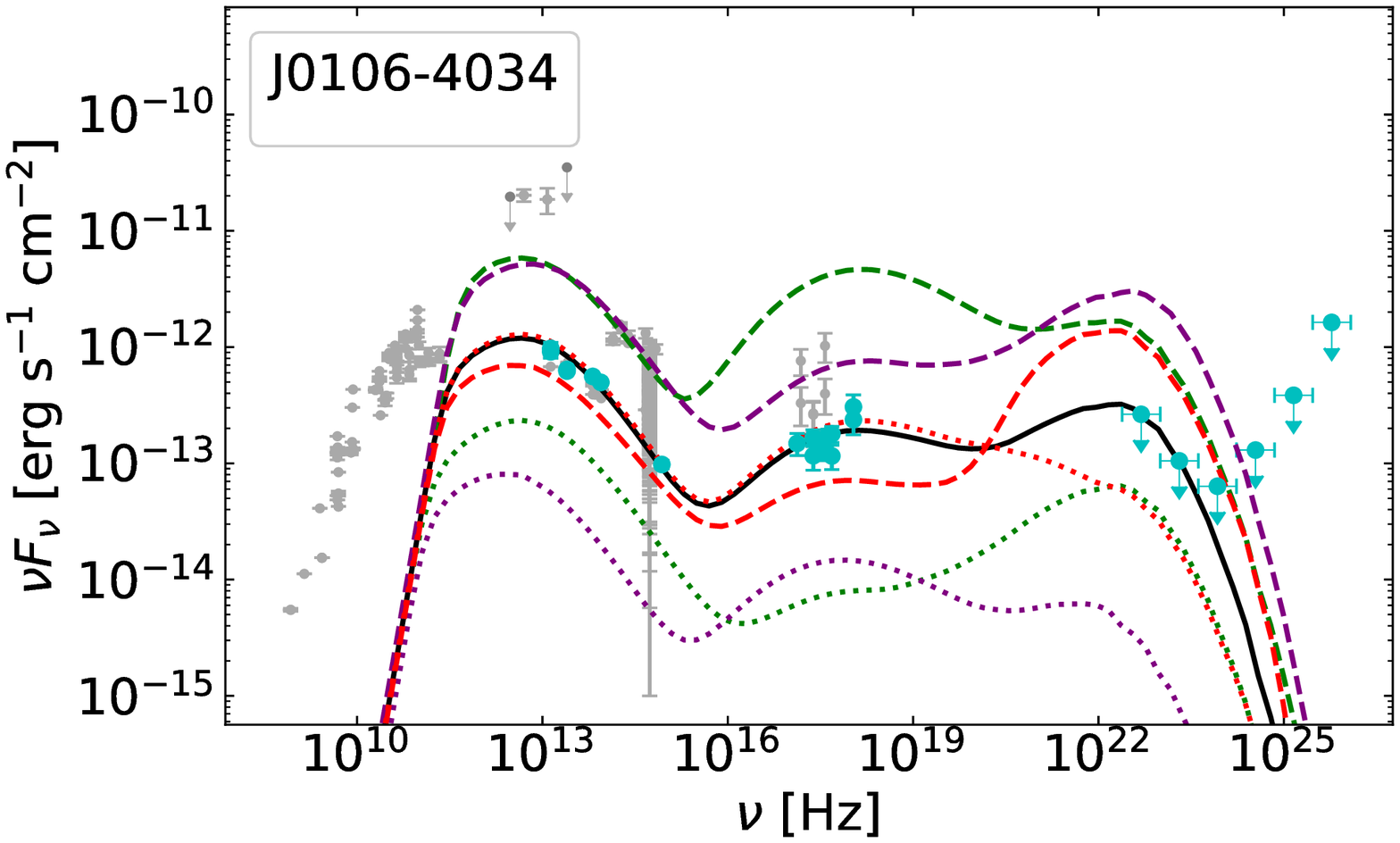}
\hspace{1.2cm}
\includegraphics[width=7cm,height=4cm]{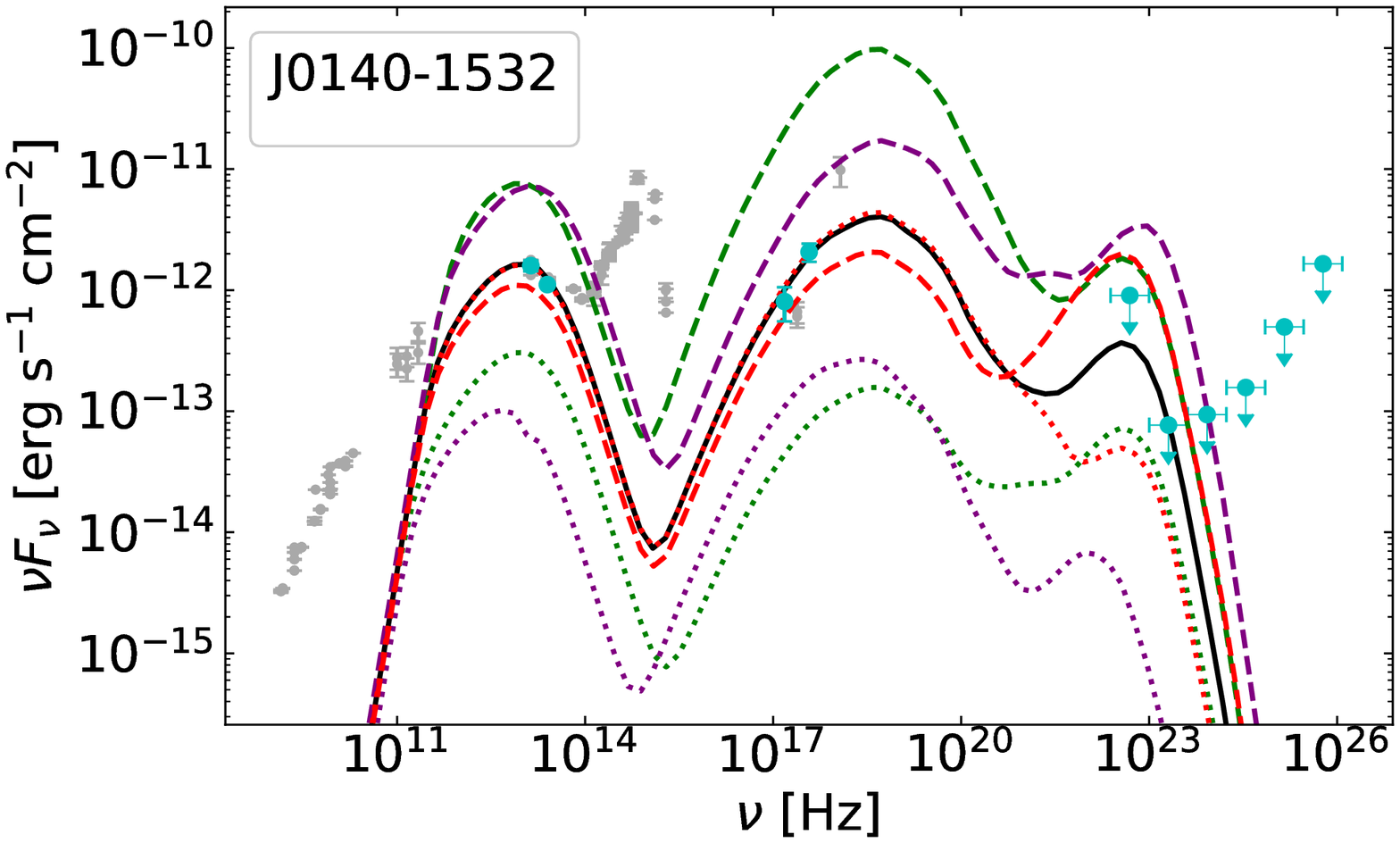}

\includegraphics[width=7cm,height=4cm]{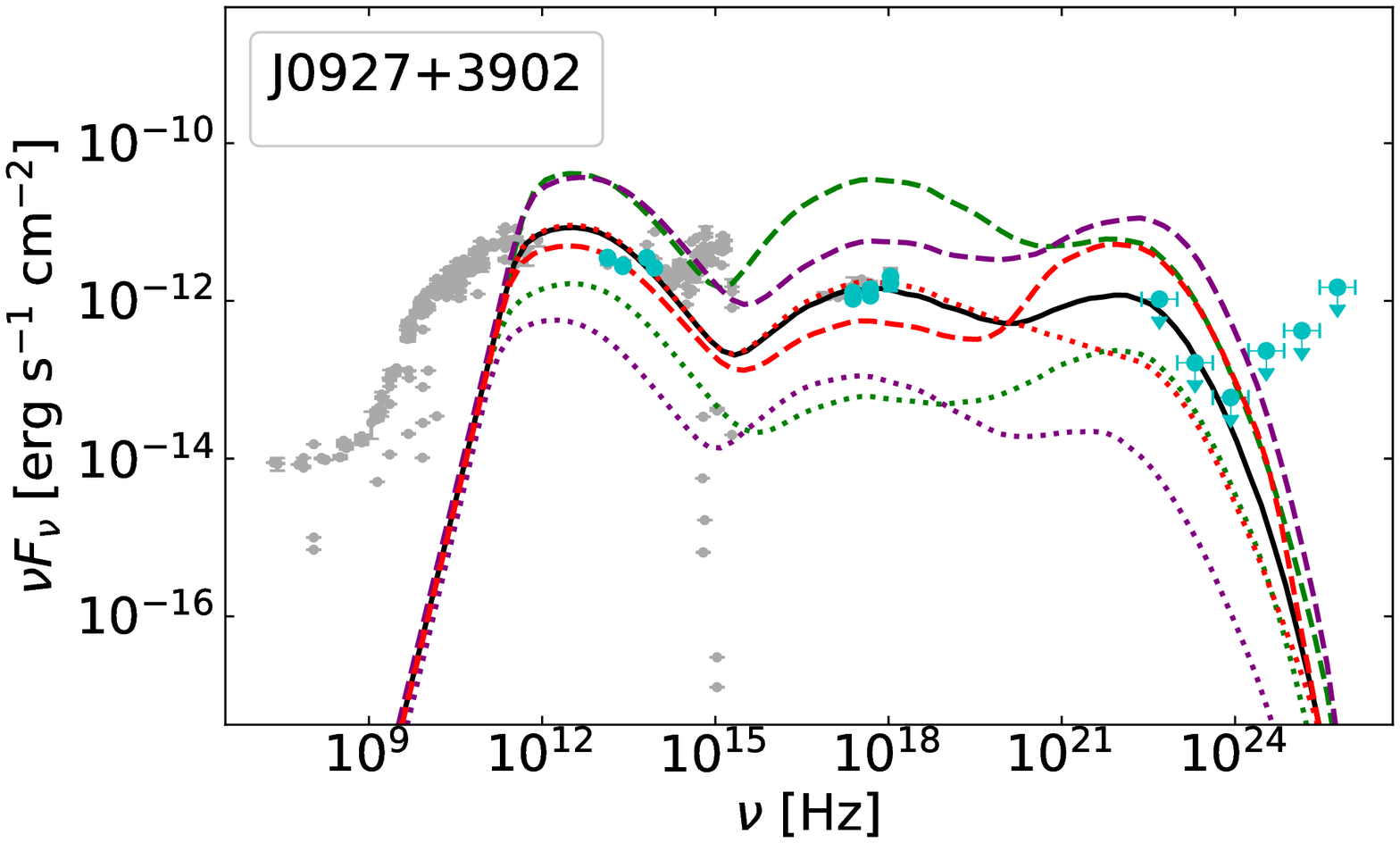}
\hspace{1.2cm}
\includegraphics[width=7cm,height=4cm]{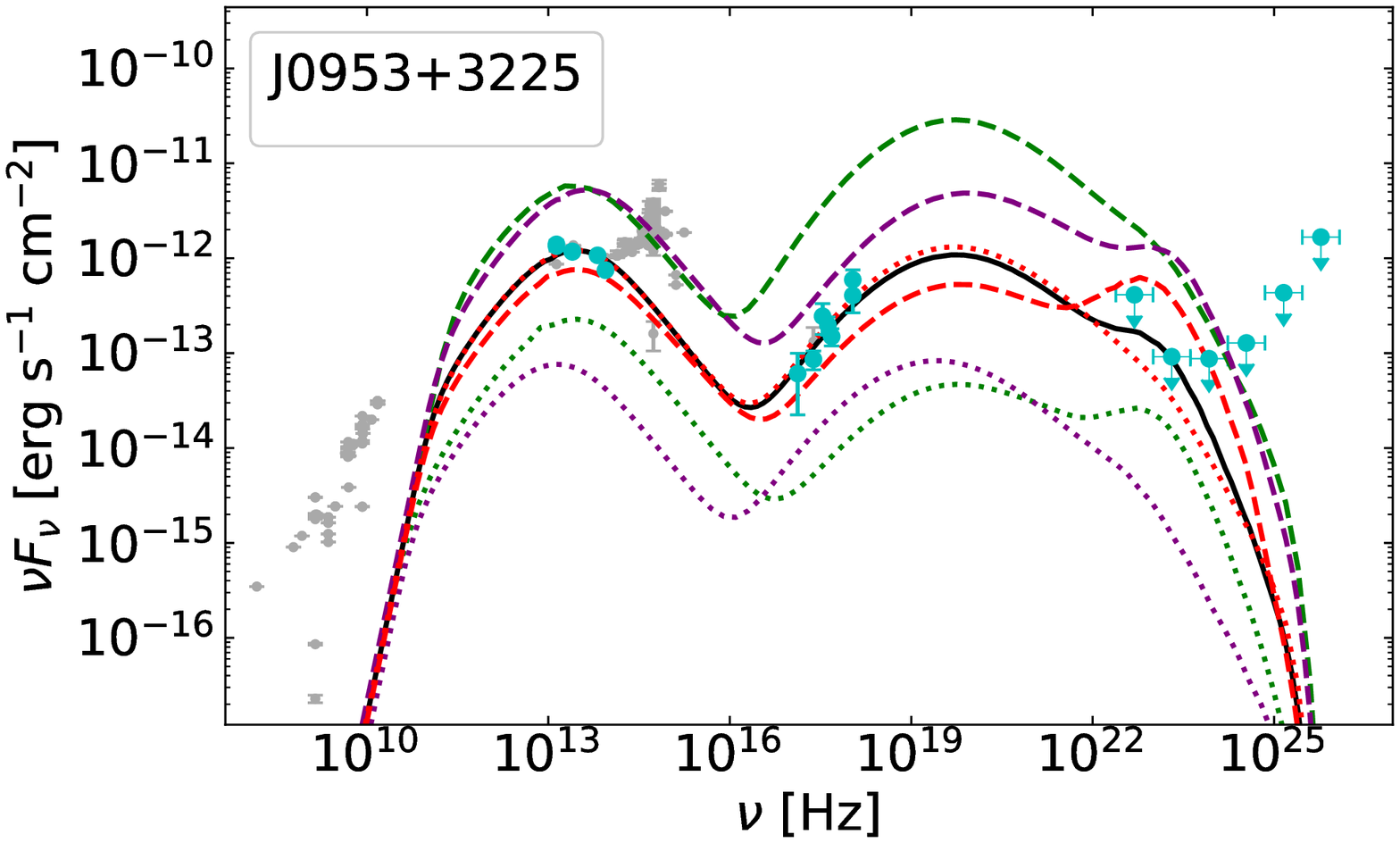}

\includegraphics[width=7cm,height=4cm]{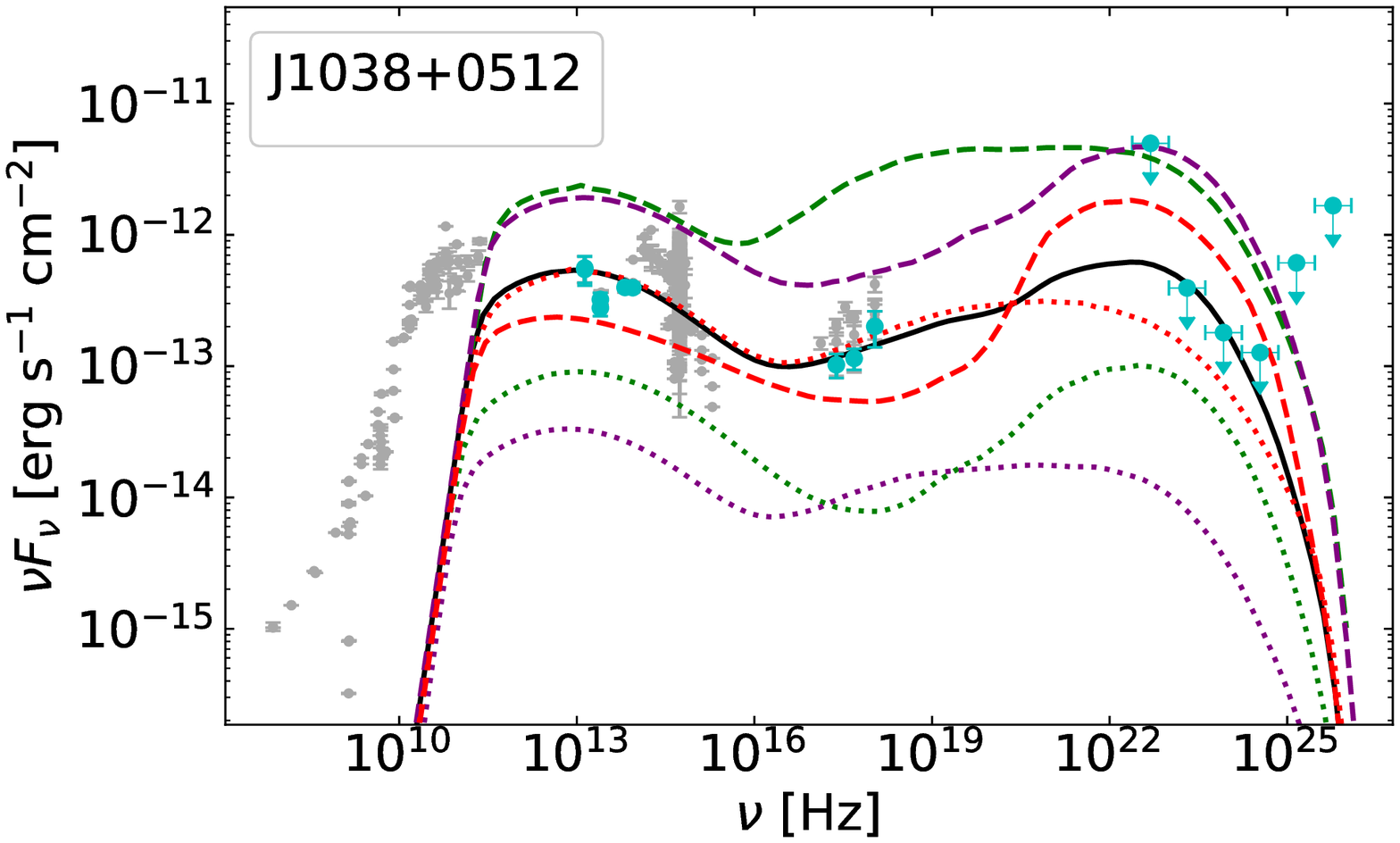}
\hspace{1.2cm}
\includegraphics[width=7cm,height=4cm]{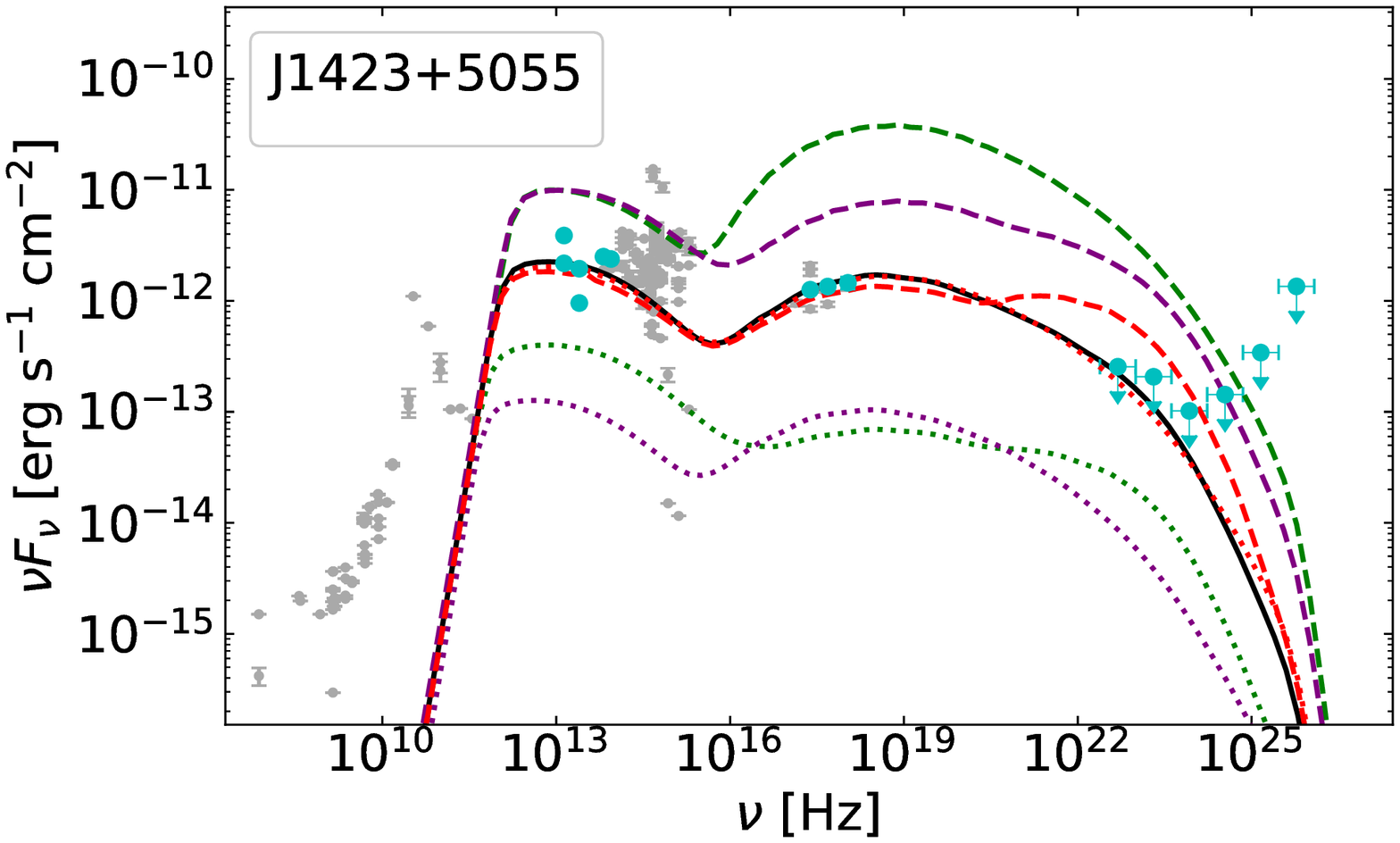}

\includegraphics[width=7cm,height=4cm]{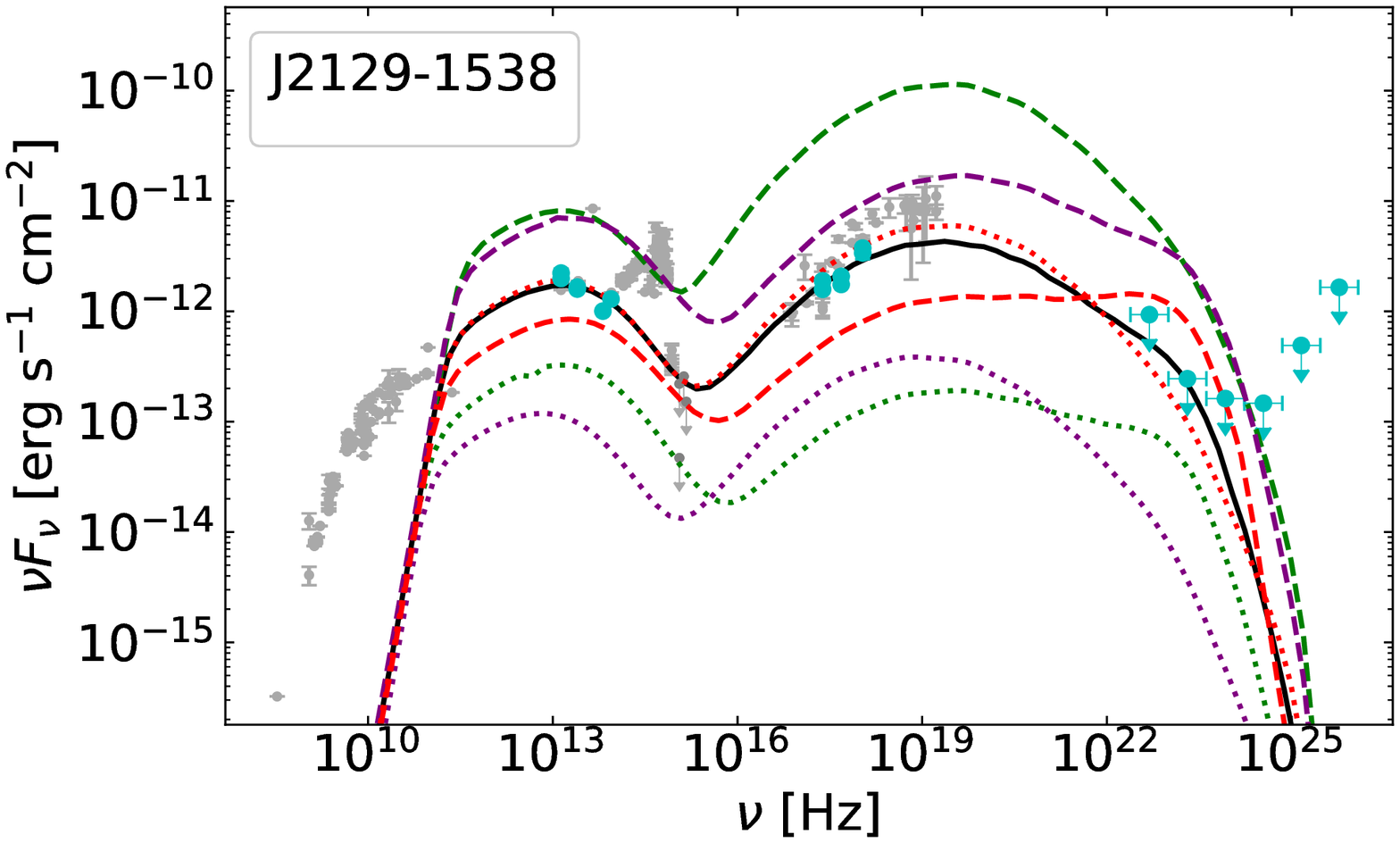}
\hspace{1.2cm}
\includegraphics[width=7cm,height=4cm]{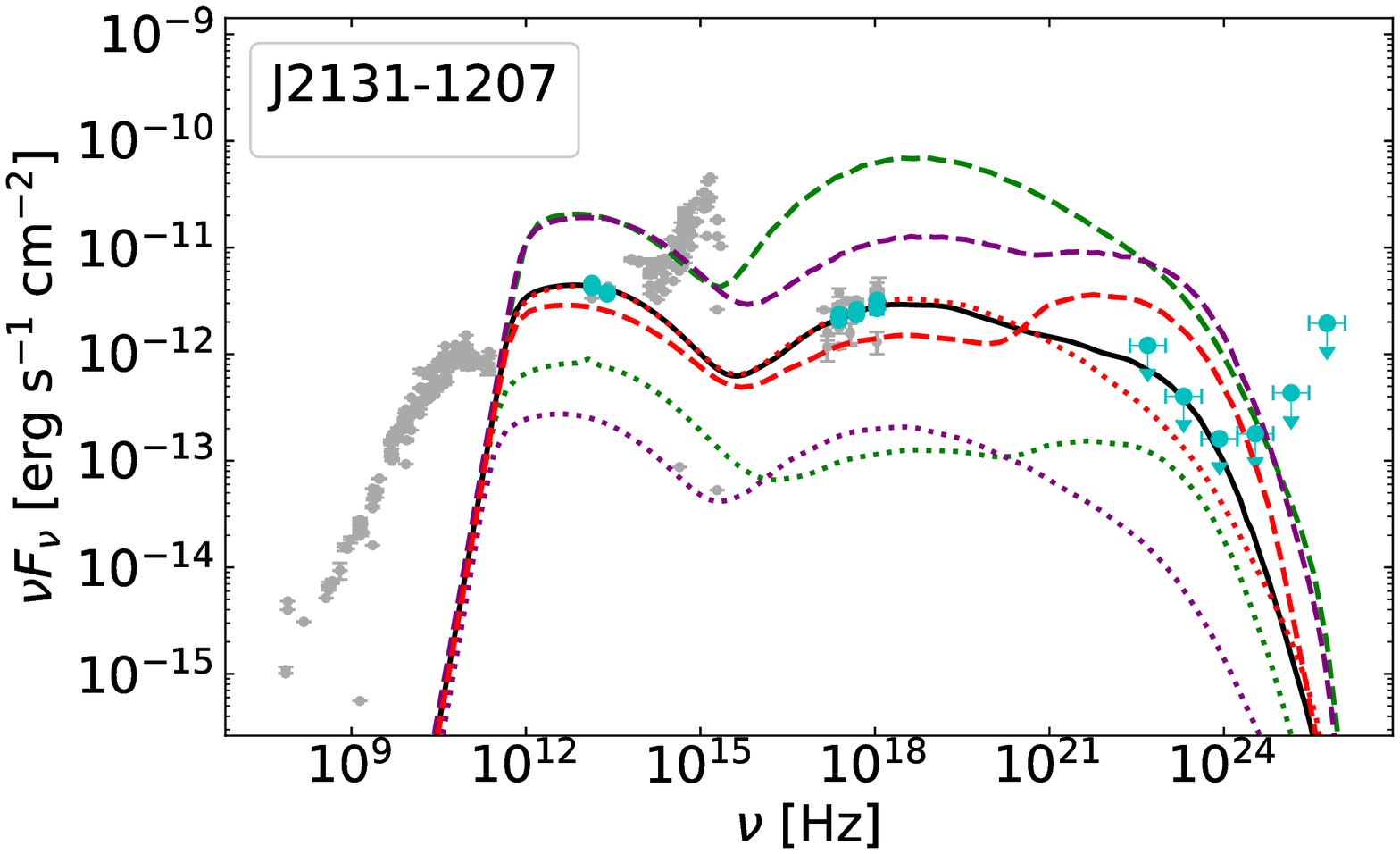}

\includegraphics[width=7cm,height=4cm]{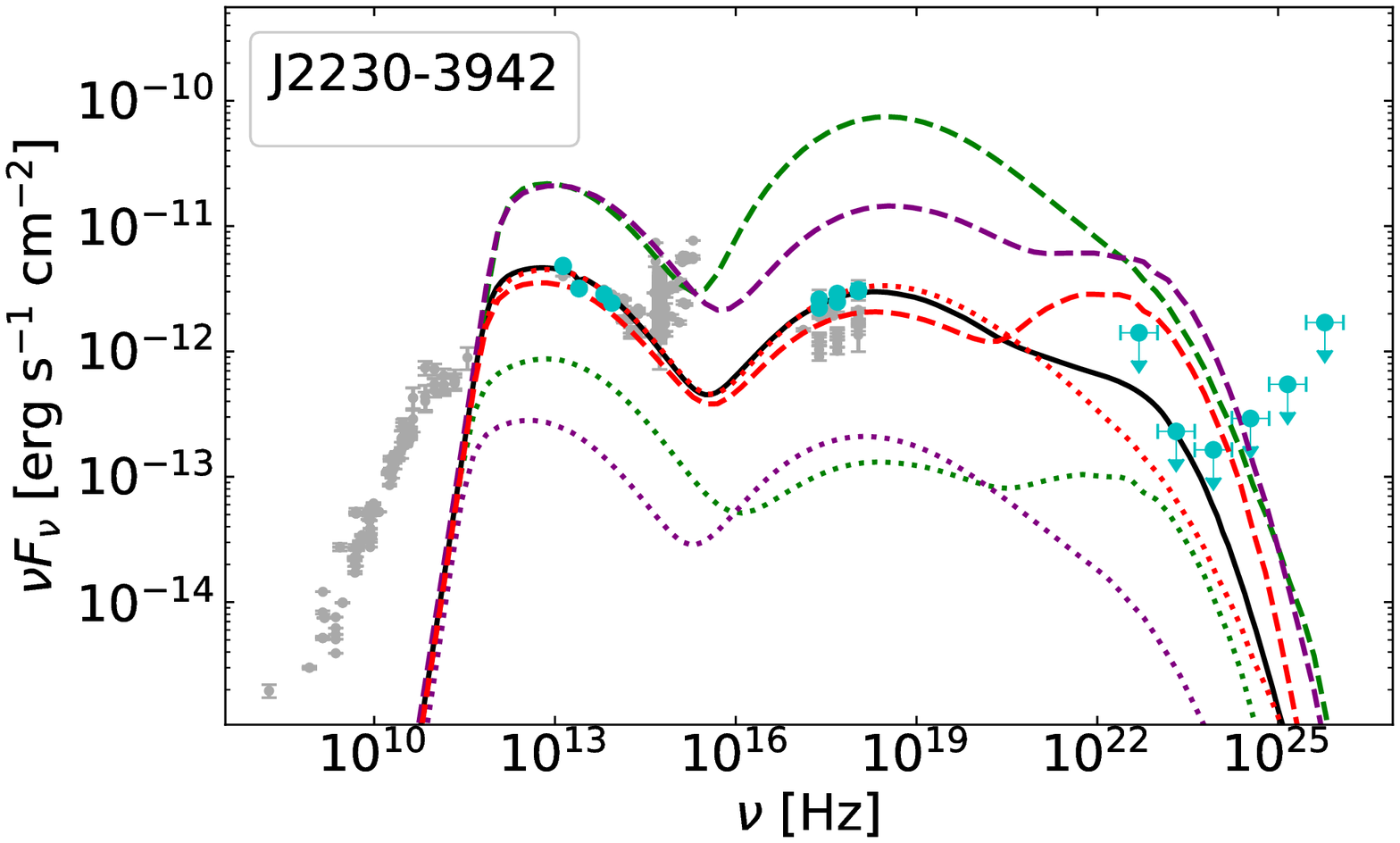}
\hspace{1.2cm}
\includegraphics[width=7cm,height=4cm]{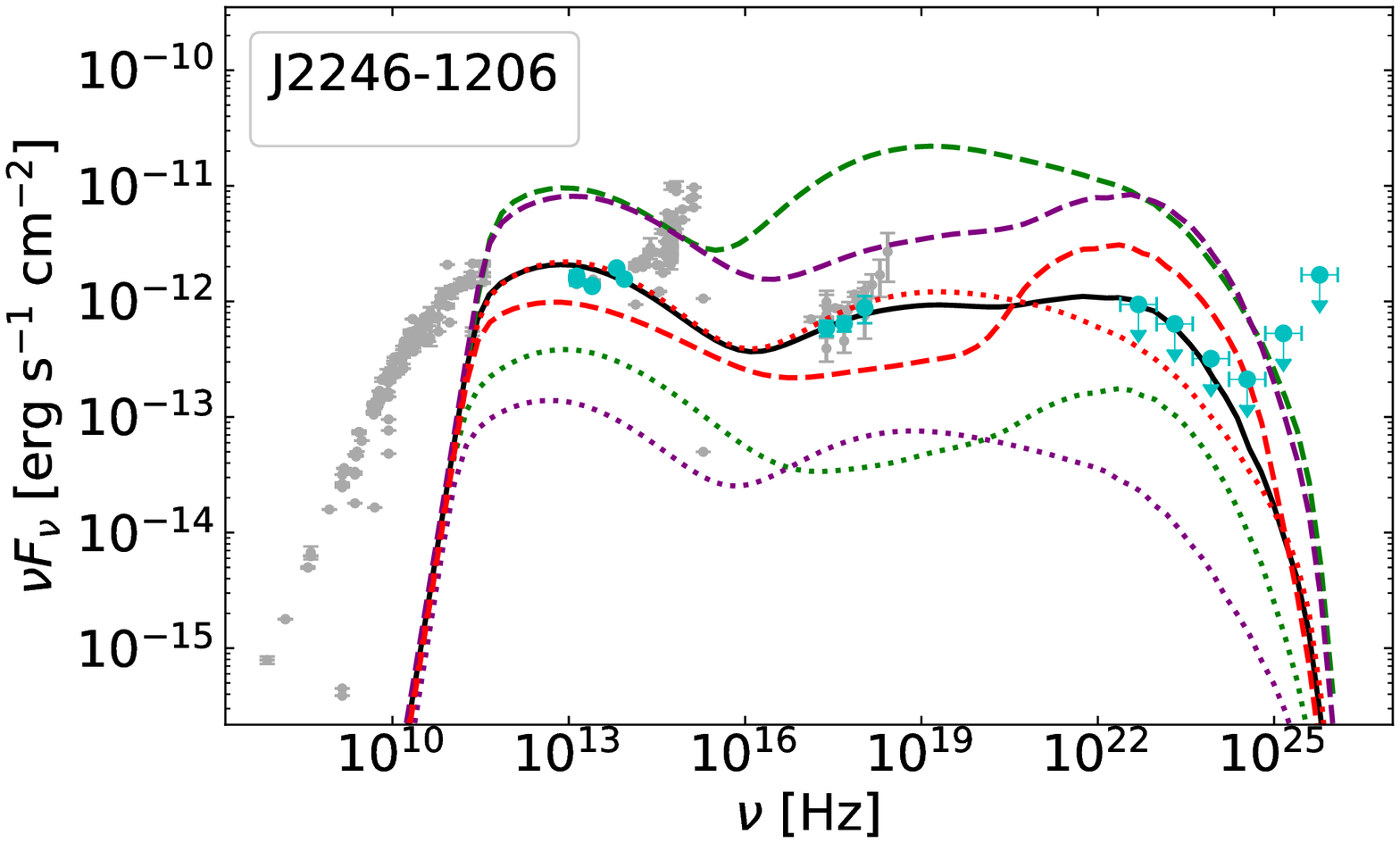}

 \caption{Modeling results of the SEDs of 11 $\gamma$-ray quiet blazars with the conventional one-zone EC model. The gray points are archival observations from SSDC. The cyan points are the quasi-simultaneous data from infrared to X-ray band, and the cyan arrows are the $Fermi$ GeV ULs. The black solid line represents the best-fit model given by the MCMC method. With the black solid lines as the standard lines, the green dashed and dotted lines are under the condition that the injected electron luminosities are increased and decreased by four times, respectively; The purple dashed and dotted lines are under the condition that Doppler factor ($\delta$=10) plus and minus 5, respectively; The red dashed and dotted lines are under the condition that the distance between dissipation region and SMBH $r$ are doubled and halved, respectively.}
 
\end{figure}
\twocolumn
\addtocounter{figure}{-1}
\begin{figure}
\centering
\includegraphics[width=7cm,height=4cm]{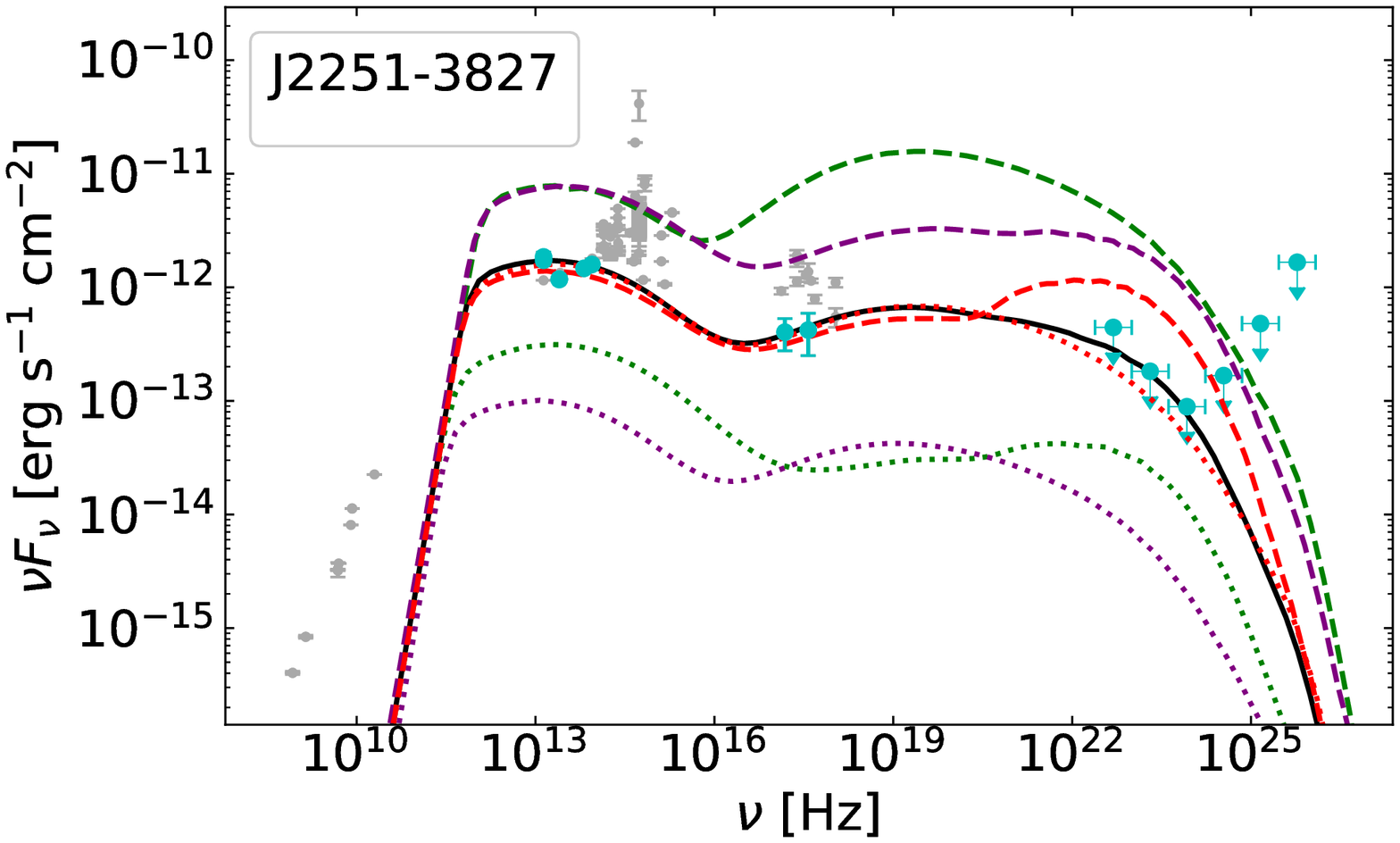}
 \caption{-$continued$}
 \label{fig1}
\end{figure}
\begin{figure}
\centering
\includegraphics[width=9cm,height=7cm]{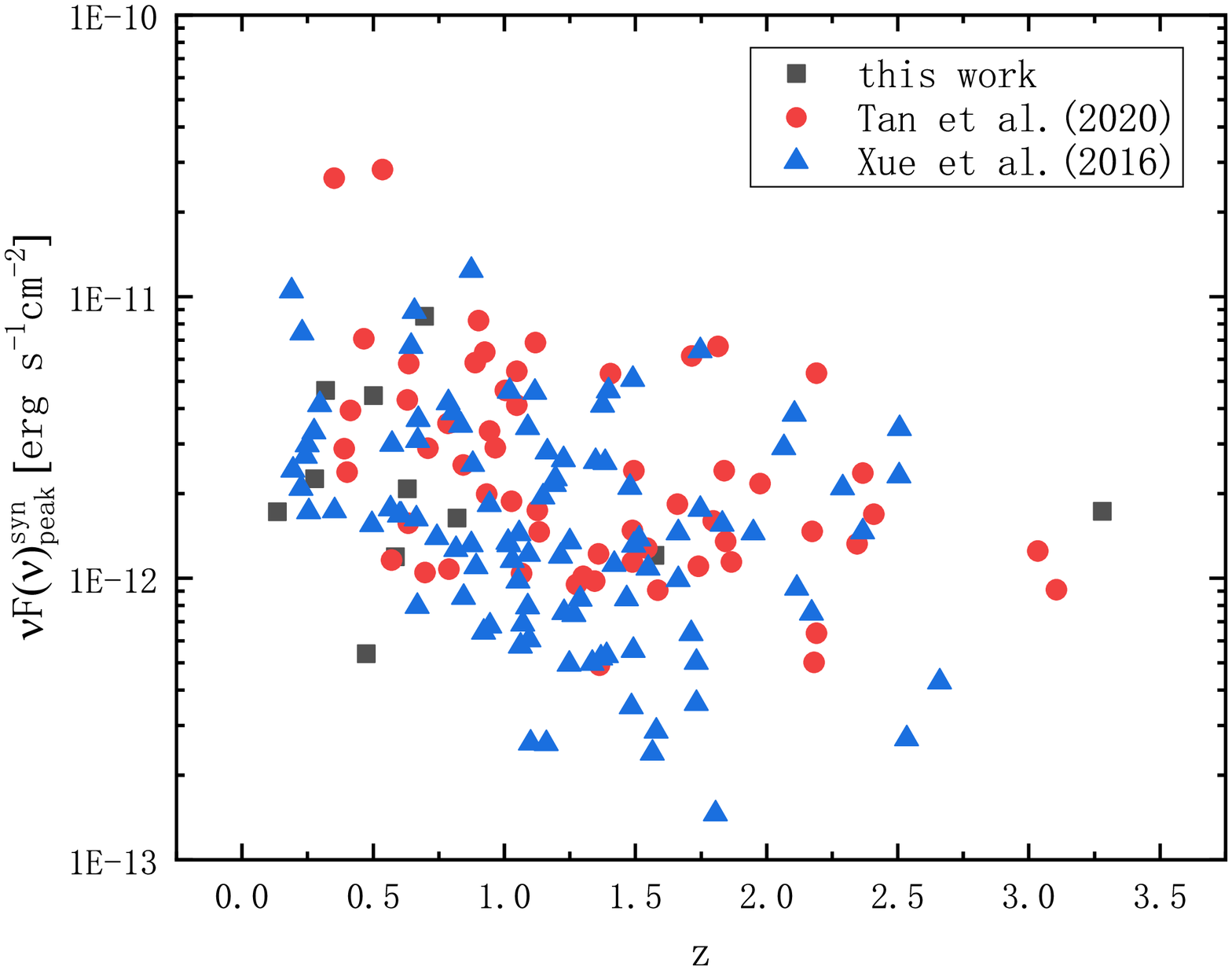}
 \caption{The comparison of the synchrotron peak flux of $\gamma$-ray quiet FSRQs between $\gamma$-ray loud FSRQs at different redshifts. The black filled squares represent the $\gamma$-ray quiet FSRQs in this work, the red filled circle and the blue filled triangle represent the $\gamma$-ray loud FSRQs from \citet{2020ApJS..248...27T} and \citet{2016MNRAS.463.3038X} respectively. }
 \label{fig2}
\end{figure}

\label{lastpage}
\end{document}